# Source-Channel Diversity for Parallel Channels

J. Nicholas Laneman, *Member, IEEE,* Emin Martinian, *Member, IEEE,*

Gregory W. Wornell, *Senior Member, IEEE,* and John G. Apostolopoulos, *Member, IEEE*

## Abstract

We consider transmitting a source across a pair of independent, non-ergodic channels with random states (e.g., slow fading channels) so as to minimize the average distortion. The general problem is unsolved. Hence, we focus on comparing two commonly used source and channel encoding systems which correspond to exploiting diversity either at the physical layer through parallel channel coding or at the application layer through multiple description source coding.

For on-off channel models, source coding diversity offers better performance. For channels with a continuous range of reception quality, we show the reverse is true. Specifically, we introduce a new figure of merit called the distortion exponent which measures how fast the average distortion decays with SNR. For continuous-state models such as additive white Gaussian noise channels with multiplicative Rayleigh fading, optimal channel coding diversity at the physical layer is more efficient than source coding diversity at the application layer in that the former achieves a better distortion exponent.

Finally, we consider a third decoding architecture: multiple description encoding with a joint source-channel decoding. We show that this architecture achieves the same distortion exponent as systems with optimal channel coding diversity for continuous-state channels, and maintains the the advantages of multiple description systems for on-off channels. Thus, the multiple description system with joint decoding achieves the best performance, from among the three architectures considered, on both continuous-state and on-off channels.

This work has been presented in part at the IEEE International Conference on Communications (ICC), Anchorage, AK, May, 2003 and the IEEE International Symposium on Information Theory (ISIT), Chicago, IL, July 2004. This work has been supported in part by Hewlett-Packard through the MIT/HP Alliance and by NSF under Grant No. CCR-9979363 as well as through NSF Graduate Research Fellowships and Oak Ridge Associated Universities (ORAU) through the Ralph E. Powe Junior Faculty Enhancement Award.

J. Nicholas Laneman is with the Department of Electrical Engineering, University of Notre Dame, Notre Dame, IN 46556. Email: jnl@nd.edu

Emin Martinian and Gregory Wornell are with the Department of Electrical Engineering and Computer Science, Massachusetts Institute of Technology, Cambridge, MA 02139. Email: {emin,gww}@allegro.mit.edu

John G. Apostolopoulos is with Hewlett-Packard Laboratories, Palo Alto, CA 94304. Email: japos@hpl.hp.com

## I. Introduction

Consider transmitting a source such as audio, video, or speech over a wireless link. Due to the nature of wireless channels, effects such as fading, shadowing, interference from other transmitters, and network congestion can cause the channel quality to fluctuate during transmission. When the channel varies on a time-scale longer than the delay constraints of the desired application, such channel fluctuations cause outages. Specifically, when the channel quality is too low, the receiver will be unable to decode the transmitted data in time to reconstruct it at the appropriate point in the source stream. Thus some frames of video or segments of speech/audio will be reconstructed at the receiver with large distortions.

As illustrated in Fig. 1, one approach to combat such channel fluctuations is to code over multiple parallel channels (*e.g.*, different frequency bands, antennas, or time slots) and leverage diversity in the channel. A variety of source and channel coding schemes can applied to this scenario, including progressive and multiple description source codes [1]–[30], broadcast channel codes [31]–[36], and hybrid analog-digital codes [37, Chapter 3] [38]–[41]; however, the best source and channel coding architecture to exploit such parallel channels is still unknown. In this paper, we examine system architectures based upon two encoding algorithms that exploit diversity in the source coding and channel coding, respectively, along with two compatible decoding algorithm for the first encoder, and one compatible decoding algorithms for the second encoder. We compare performance of these systems by studying their average distortion performance on a various block fading channel models.

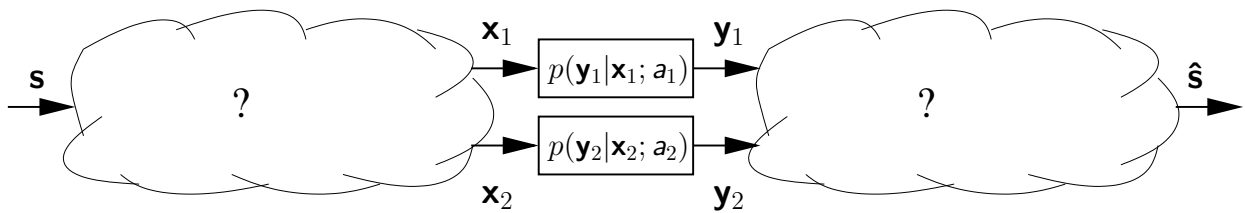

Fig. 1. Conceptual illustration of the parallel diversity coding problem considered in this paper. An encoder must map a source sequence, $\mathbf{s}$, into a pair of channel inputs $\mathbf{x}_1$ and $\mathbf{x}_2$ *without* knowing the channel states $a_1$ and $a_2$. A decoder must map the channel outputs $\mathbf{y}_1$ and $\mathbf{y}_2$ along with knowledge of the channel states into an estimate of the source, $\hat{\mathbf{s}}$. The optimal encoding and decoding architecture is unknown.

More specifically, Fig. 2 illustrates the two classes of encoders we consider. In the *channel coding diversity* system of Fig. 2(a), the source $s$ is encoded into $\hat{s}$ by a single description (SD) source coder. Next $\hat{s}$ is jointly encoded into $(x_1, x_2)$ by the channel coder and transmitted across a parallel channel. For the

Decoder

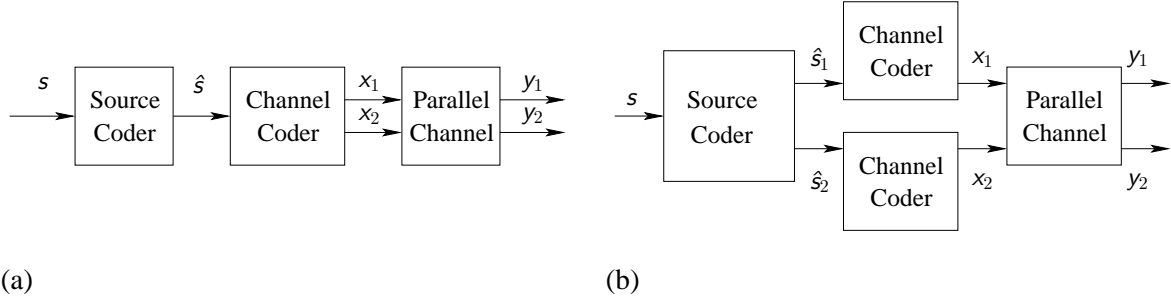

Fig. 2.   Block diagrams for (a) channel coding diversity and (b) source coding diversity.

*source coding diversity* system of Fig. 2(b), the source $s$ is encoded into $\hat{s}_1$ and $\hat{s}_2$ by a multiple description (MD) source coder. Each $\hat{s}_i$ is then separately encoded into $x_i$ by a channel coder and transmitted across the appropriate channel.

Since the encoders in Fig. 2 exploit the inherent diversity of a parallel channel in qualitatively different ways, we focus on the following two questions:

1) Which of the basic architectures in Fig. 2 achieves the smallest average distortion? If neither architecture is universally best, for what channels is one architecture better than the other?

2) Is there a way to combine the best features of both systems in Fig. 2?

Essentially, the answers we develop can be illustrated through Fig. 3. For channel coding diversity, the source codeword, $\hat{s}$, can be reliably decoded only if the *total* channel quality is high enough to support the transmission rate. So this system achieves diversity in the sense that even if one of the channels is bad, then as long as the overall channel quality is good, the receiver will still be able to recover the encoded source. In contrast, for source coding diversity, each source codeword $\hat{s}_i$ can be decoded if the quality of the corresponding *individual* channel is high enough. This system achieves diversity in the sense that even if one of the channels is bad and one description is unrecoverable, then as long as the other channel is good and the remaining description is recovered, a low fidelity source reconstruction is obtained. If both channels are good and both descriptions are successfully decoded, then they are combined to form a high fidelity reconstruction.

Fig. 3 compares the two systems when the source coders are designed to achieve the same distortion if all source codewords are successfully decoded (*i.e.*, in region III). Furthermore, in region I, both systems fail to decode and again have the same distortion. In regions II and V, channel coding diversity is superior since the channel conditions are such that at most one source codeword is decoded under source coding diversity. Conversely, in region IV, source coding diversity is superior since one source codeword is received, and channel coding diversity fails to decode. Therefore our first question about which of the architectures in Fig. 2 is best, is essentially a question about which region the channel quality is most likely to lie in. If regions II and V are more probable, channel coding diversity will be superior; conversely, if regions IV are more likely, source coding diversity will be superior.

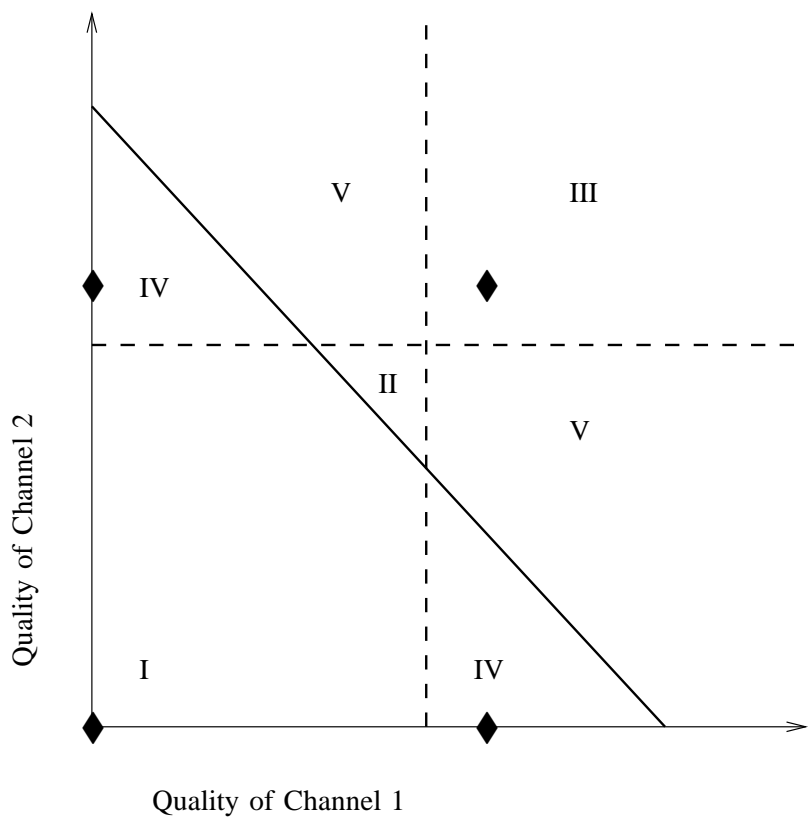

Fig. 3. Conceptual illustration of successful decoding regions for source and channel coding diversity systems designed to have the same distortion when all codewords are received. For channel coding diversity, the receiver will be able to decode the transmitted source description if the sum of the channel quality exceeds a threshold represented by the solid diagonal line. For source coding diversity, the first (respectively, second) source description will be successfully decoded provided the first (resp., second) channel quality exceeds the vertical (resp., horizontal) dashed line. The ◆'s represent the four possible channel qualities for a packet loss channel where each channel is either on or off.

As a specific example, in the classic MD coding problem modeling link failure or packet erasure [28], each channel is either off, in which case no information can be communicated, or supports a particular rate. The four channel conditions for this scenario are indicated by ◆'s in Fig. 3 for an example packet erasure channel. For such discrete models, source coding diversity is clearly superior, since both SD and MD source coding achieve the same distortions in regions I and III, but channel coding diversity fails completely in region IV. In this region, source coding diversity recovers one source codeword and produces a low fidelity reconstruction of the source.

The opposite occurs for channels where a continuous range of rates can potentially be supported (*e.g.*, additive white Gaussian noise channels with Rayleigh fading). For these channels, the channel quality is essentially more likely to lie in region II than in IV and thus channel coding diversity is superior. Specifically, we characterize performance by analyzing how quickly the average distortion decays as a function of the signal-to-noise ratio (SNR) for various systems. We refer to the slope of the distortion versus SNR on a log-log plot as the "distortion exponent" and use this as our figure of merit. In particular, our analysis shows that optimal channel coding diversity is generally superior to source coding diversity on continuous channels in the sense that an optimal channel coding diversity architecture achieves a better distortion exponent than a source coding diversity architecture.

Since source coding diversity is best for on-off channels, and optimal channel coding diversity is best for continuous state channels, our second question of whether there exists an architecture that combines the advantages of both becomes relevant. In addition to our analysis of the two previously known diversity architectures in Fig. 2, our second main contribution is the description of a new joint source-channel decoding architecture which achieves the best qualities of both. Specifically, to perform well on both continuous state channels and on-off channels we do not propose a third *encoding* architecture, but a third new *joint decoding* architecture. We show that the main inefficiency of source coding diversity on continuous state channels results from the channel decoders ignoring the correlation between the multiple descriptions. By explicitly accounting for the structure of the source encoding when performing channel decoding, we prove a coding theorem characterizing the performance of source coding diversity with joint decoding. We show that such a system can achieve the same performance as optimal channel coding diversity on continuous channels and the same performance as source coding diversity for on-off channels.

### A. Related Research

The problem of MD coding was initially studied from a rate-distortion perspective, having been formalized by Gersho, Witsenhausen, Wolf, Wyner, Ziv, and Ozarow at the 1979 IEEE Information Theory Workshop. Their initial contributions to the problem appear in [29], [42]–[44]. El Gamal & Cover develop an achievable rate region for two descriptions in [28], and this region is shown to be optimal for the Gaussian source, with mean-square distortion, by Ozarow [44]. Specialized results for the binary symmetric source, with Hamming distortion, are developed by Berger & Zhang [24], [26], [45] and Ahlswede [27]. Zamir [23] develops high-rate bounds for memoryless sources. Most recently, work by Venkatarami *et. al* [3], [21] provides achievable rate regions for many descriptions that generalize the results in [26], [28]. Important special cases of the MD coding problem have also been examined, including successive refinement, or layered coding, [1], [46] and certain symmetric cases [2], [20].

Some practical approaches to MD coding include MD scalar quantization, dithered MD lattice quantization, and MD transform coding. Vaishampayan [25] pioneered the former, Frank-Dayan and Zamir considered the use of dither [7], and Wang, Orchard, Vaishampayan, and Reibman [22] and later Goyal & Kovacevic [16] studied the latter. See [17] for a thorough review of both approaches. Recently, the design of MD video coders has received considerable attention [4], [8]–[10], [13], [19]

All of the classical work on MD coding utilizes an "on-off" model for the channels or networks under consideration, without imposing strict delay constraints. More specifically, source codes are designed assuming that each description is completely available (error-free) at the receiver, or otherwise completely lost. Furthermore, the likelihood of these events occurring is independent of the choice of source coding rates. Under such conditions, it is not surprising that MD coding outperforms SD coding; however, for many practical channel and network environments, these conditions do not hold. For example, in delay constrained situations, suitable for real-time or interactive communication, descriptions may have to be encoded as multiple packets, each of which might be received or lost individually. Furthermore, congestion and outage conditions often depend heavily upon the transmission rate. Thus, it is important to consider MD coding over more practical channel models, as well as to fairly compare performance with SD coding.

Some scattered work is appearing in this area. Ephremides *et. al* [11] examine MD coding over a parallel queue channel, compare to SD coding, and show that MD coding offers significant advantages under high traffic (congestion) situations. This essentially results because the MD packets are more compact than SD packets, and indicates the importance of considering the influence of rate on congestion. Coward *et. al* [6], [15] examine MD coding over several channel models, including memoryless symbol-erasure and symbol-error channels, as well as block fading channels. For strict delay constraints, they show that MD outperforms SD; for longer delay constraints, allowing for more sophisticated channel coding, they show that SD outperforms MD. Thus, the impact of delay constraints are important. This paper examines fading conditions similar to those in [6], [15], but considers a wider variety of channel coding and decoding options, with an emphasis on architectural considerations as well as performance.

### B. Outline

We begin by summarizing our system model in Section II. Section III studies on-off channels, Section IV treats continuous state channels, and Section V develops source coding diversity with joint decoding. Many of the more detailed proofs are deferred to Appendices. Finally, Section VI closes the paper with some concluding remarks and directions for further research.

## II. System Model

Fig. 1 depicts the general system model we consider in this paper. Our objective is to design and evaluate methods for communicating a source signal $s$ with small distortion over certain channels with independent parallel components. In particular, focusing on memoryless source models for simplicity of exposition, we consider non-ergodic channels models in which delay constraints or limited channel variations limit the effective blocklength at the encoder. Of many possible examples, we focus on on-off channels and additive noise channels with block fading. While cross-layer design is generally acknowledged to yield superior performance to layered design, simultaneously optimizing all facets of a system is usually too complex. Hence we consider various architectures based upon using a classical system at one layer combined with an optimized system at another layer. In the remainder of this section, after briefly introducing some notation, we summarize the source and channel models, discuss architectural options for encoding and decoding, and review high-resolutions approximations for the various source coding algorithms employed throughout the paper.

### A. Notation

Vectors and sequences are denoted in bold (*e.g.*, $\mathbf{x}$) with the $i$th element denoted as $x[i]$. Random variables are denoted using the sans serif font (*e.g.*, $\mathsf{x}$) while random vectors and sequences are denoted with bold sans serif (*e.g.*, $\mathbf{\mathsf{x}}$). We denote mutual information, differential entropy, and expectation as $I(\mathsf{x}; \mathsf{y})$, $h(\mathsf{x})$, $E[\mathsf{x}]$. Calligraphic letters denote sets (*e.g.*, $s \in \mathcal{S}$). When its argument is a set or alphabet, $|\cdot|$ denotes the cardinality of the argument. To simplify the discussion of architectures, we use the symbols $\mathrm{ENC}(\cdot)$ and $\mathrm{DEC}(\cdot)$ to denote a generic encoder and decoder. To specialize this generic notation to one of the architectures discussed in Section II-D, we will employ subscripts representing the relevant system variables.

### B. Source Model

We model the source as a sequence of independent and identically distributed (i.i.d.) samples $s[k]$. For example, such a discrete-time source may be obtained from sampling a continuous-time, appropriately band-limited, white-noise random process. We denote the probability density for the discrete-time source sequence $s[k]$ as

$$p_{\mathbf{s}}(\mathbf{s}) = \prod_{k=1}^{K} p_s(s[k]) \ . \tag{1}$$

We assume that the process is such that the differential entropy, $h(s)$, and second moment, $\mathrm{E}[s^2]$, both exist and are finite.

To measure quality of the communication system, we employ a distortion measure between the source signal $s$ and its reconstruction $\hat{s} \in \hat{\mathcal{S}}$. Specifically, given a per-letter distortion measure $d(s[k], \hat{s}[k])$, we extend it additively to blocks of source samples, *i.e.*,

$$d(\mathbf{s}, \hat{\mathbf{s}}) = \sum_{k=1}^{K} d(s[k], \hat{s}[k]) \ . \tag{2}$$

We may characterize performance in terms of various statistics of the distortion, viewed as a random variable. In particular, we focus on the expected distortion

$$D = \mathrm{E}[d(\mathbf{s}, \hat{\mathbf{s}})] \ . \tag{3}$$

Throughout our development, we will emphasize squared-error distortion, for which $d(s, \hat{s}) = (s - \hat{s})^2$; in this case, (3) is the mean-square distortion.

### C. (Parallel) Channel Model

The channel depicted by Fig. 1 consists of two branches, each of which corresponds to an independent channel with independent states. Specifically, a channel input block, $\mathbf{x}$, consists of two sub-blocks, $\mathbf{x}_1$ and $\mathbf{x}_2$, and the corresponding channel output block, $\mathbf{y}$, consists of the two sub-blocks, $\mathbf{y}_1$ and $\mathbf{y}_2$. The channel states are denoted by random variables $a_1$ and $a_2$, respectively. The channel law is the product of the two independent sub-channel laws:

$$p_{\mathbf{y}_1, \mathbf{y}_2, a_1, a_2 | \mathbf{x}_1, \mathbf{x}_2}(\mathbf{y}_1, \mathbf{y}_2, a_1, a_2 | \mathbf{x}_1, \mathbf{x}_2) = p_{\mathbf{y}, a | \mathbf{x}}(\mathbf{y}_1, a_1 | \mathbf{x}_1) \cdot p_{\mathbf{y}, a | \mathbf{x}}(\mathbf{y}_2, a_2 | \mathbf{x}_2) =$$

$$p_a(a_1) \cdot p_a(a_2) \prod_{i=1}^{n_c} \left[ p_{y|x,a}(y_1[i]|x_1[i], a_1) \cdot p_{y|x,a}(y_2[i]|x_2[i], a_2) \right] \ . \tag{4}$$

For simplicity, we only consider channels for which the input distribution that maximizes the mutual information is independent of the channel state. Throughout the paper we consider the case where both the transmitter and receiver know the channel state distribution $p_a$ and the channel law $p_{y|x}$, but only the receiver knows the realized channel states and channel outputs.

To examine fundamental performance and compare between systems, we analyze random coding over these non-ergodic channels using outage probability [47] as a performance measure. Briefly, because the mutual information $I$, corresponding to the supportable transmission rate of the channel, is a function of the fading coefficients or other channel uncertainty, it too is a random variable. For fixed transmission rate $R$ (in nats/channel use), the outage probability $\Pr[I < R]$ measures channel coding robustness to uncertainty in the channel.[1]

The structure of the channel coding and decoding affects the form of the outage probability expression [47]. If coding is performed over only the first component channel, then the probability of decoding failure is $\Pr[I(\mathbf{x}_1; \mathbf{y}_1) < R]$. If repetition coding is performed across the parallel channels, then a single message is encoded as $\mathbf{x}_1 = \mathbf{x}_2 = \mathbf{x}$. With selection combining at the receiver, the probability of decoding failure is $\Pr\{\max[I(\mathbf{x}; \mathbf{y}_1), I(\mathbf{x}; \mathbf{y}_2)] < R\}$; with optimal maximum-ratio combining at the receiver, the probability of decoding failure is $\Pr\{I(\mathbf{x}; \mathbf{y}_1, \mathbf{y}_2) < R\}$. Finally, if optimal parallel channel coding is performed using a pair of jointly-designed codebooks with $\mathbf{x}_1$ and $\mathbf{x}_2$ independent, the probability of decoding failure is $\Pr[I(\mathbf{x}_1; \mathbf{y}_1) + I(\mathbf{x}_2; \mathbf{y}_2) < R]$.

### D. Architectural Options

In this section, we specify some architectural options for encoding and decoding in the source-channel diversity system depicted in Fig. 1.

*1) Joint Source-Channel Diversity:* In the most general setup, joint source-channel diversity consists of a pair of mappings $(\text{ENC}_{\mathbf{x}_1, \mathbf{x}_2 \leftarrow \mathbf{s}}, \text{DEC}_{\hat{\mathbf{s}} \leftarrow \mathbf{y}_1, \mathbf{y}_2})$. The encoder $\text{ENC}_{\mathbf{x}_1, \mathbf{x}_2 \leftarrow \mathbf{s}}$ maps a sequence of $K$ source letters into $N$ pairs of channel inputs; correspondingly, the decoder maps $N$ pairs of channel outputs into $K$ reconstruction letters. The ratio $N/K$ (sometimes referred to as the processing gain, excess bandwidth, or bandwidth expansion factor) is denoted with the symbol $\beta \triangleq N/K$.[2] Mathematically,

$$\text{ENC}_{\mathbf{x}_1, \mathbf{x}_2 \leftarrow \mathbf{s}} : \mathcal{S}^K \longrightarrow \mathcal{X}_1^N \times \mathcal{X}_2^N \tag{5}$$

$$\text{DEC}_{\hat{\mathbf{s}} \leftarrow \mathbf{y}_1, \mathbf{y}_2} : \mathcal{Y}_1^N \times \mathcal{Y}_2^N \longrightarrow \hat{\mathcal{S}}^K \ . \tag{6}$$

---

[1]Mutual information is often used to measure channel robustness when long block lengths are allowed. In [48], however, Zheng and Tse show that mutual information (viewed as a random variable), and more specifically outage probability, is a relevant quantity for finite block lengths since outage probability dominates error probability. This suggests that outage can be a relevant quantity even for very tight delay constraints at high SNR.

[2]The bandwidth expansion ratio in [49] (denoted by $L$) is defined slightly differently from $\beta$. Specifically, since [49] considers a complex source and Rayleigh fading Gaussian noise channel, $L = 2\beta$.

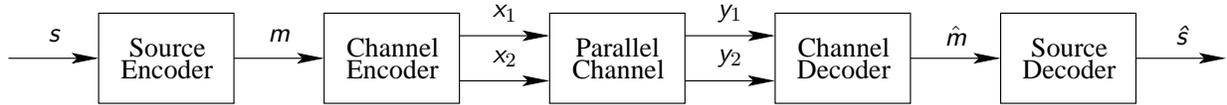

Fig. 4. Channel coding diversity.

If the image of $\mathrm{ENC}_{\mathbf{x}_1,\mathbf{x}_2\leftarrow\mathbf{s}}$, *i.e.*, $\mathrm{ENC}_{\mathbf{x}_1,\mathbf{x}_2\leftarrow\mathbf{s}}(\mathcal{S}^K)$, is finite, we define the *rate* of the code as

$$R = \frac{\ln|\mathrm{ENC}_{\mathbf{x}_1,\mathbf{x}_2\leftarrow\mathbf{s}}(\mathcal{S}^K)|}{N} \ , \tag{7}$$

which has units of nats per parallel channel use.

Regarding the non-ergodic nature of the channels, we consider situations in which $K$ is large enough to average over source fluctuations, *i.e.*, the source is ergodic, but $N$ is not large enough to average over channel variations, *i.e.*, the channel is non-ergodic.

*2) Channel Coding Diversity:* From one perspective, a natural way to exploit diversity in the channel is to employ repetition or more powerful channel codes applied to a single digital representation of the source. In such scenarios, Fig. 1 specializes to that shown in Fig. 4. Such channel coding diversity consists of a source pair of encoder and decoder mappings $(\mathrm{ENC}_{m\leftarrow\mathbf{s}}, \mathrm{DEC}_{\hat{\mathbf{s}}\leftarrow\hat{m}})$ and a channel pair of encoder and decoder mappings $(\mathrm{ENC}_{\mathbf{x}\leftarrow m}, \mathrm{DEC}_{\hat{m}\leftarrow\mathbf{y}})$. As in classical rate-distortion source coding, the source encoder maps a sequence of $K$ input letters to a finite index, and the source decoder maps an index into a sequence of $K$ reconstruction letters:

$$\mathrm{ENC}_{m\leftarrow\mathbf{s}} : \mathcal{S}^K \longrightarrow \{1, 2, \ldots, |\mathcal{M}|\} \tag{8}$$

$$\mathrm{DEC}_{\hat{\mathbf{s}}\leftarrow\hat{m}} : \{0, 1, 2, \ldots, |\mathcal{M}|\} \longrightarrow \hat{\mathcal{S}}^K \tag{9}$$

Further, as in classical channel coding, the channel encoder maps an index into $N$ pairs of channel inputs, and the channel decoder maps $N$ pairs of channel outputs into an index:

$$\mathrm{ENC}_{\mathbf{x}\leftarrow m} : \{1, 2, \ldots, |\mathcal{M}|\} \longrightarrow \mathcal{X}_1^N \times \mathcal{X}_2^N \tag{10}$$

$$\mathrm{DEC}_{\hat{m}\leftarrow\mathbf{y}} : \mathcal{Y}_1^N \times \mathcal{Y}_2^N \longrightarrow \{0, 1, \ldots, |\mathcal{M}|\} \ . \tag{11}$$

Note that we include the index $0$ at the output of the channel decoder and input to the source decoder. This serves as a flag in the event of a (detected) channel coding error or outage in which case the source decoder reconstructs to the mean of the source.

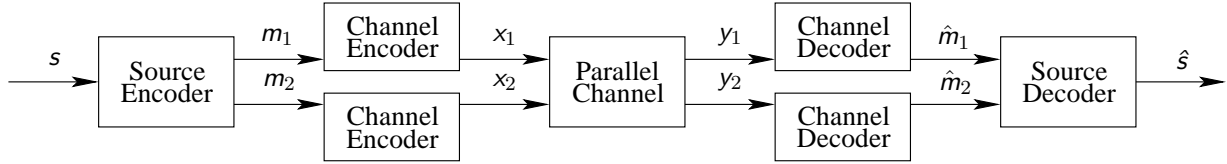

Fig. 5.   Source coding diversity system model described more precisely in Section II-D.3.

For the channel coding diversity approach, a key parameter is the rate defined by

$$R = \frac{\ln |\mathcal{M}|}{N} \ , \tag{12}$$

where again the units are nats per parallel channel use.

*3) Source Coding Diversity:*  Instead of exploiting diversity through channel coding, an emerging class of source coding algorithms based upon MD coding allows diversity to be exploited by the source coding layer.

For such source coding diversity, the block diagram of Fig. 1 specializes to that shown in Fig. 5. Source coding diversity employs two independent, but otherwise classical, channel encoder and decoder pairs $(\text{ENC}_{\mathbf{x}_1 \leftarrow m_1}, \text{DEC}_{\hat{m}_1 \leftarrow \mathbf{y}_1})$ and $(\text{ENC}_{\mathbf{x}_2 \leftarrow m_2}, \text{DEC}_{\hat{m}_2 \leftarrow \mathbf{y}_2})$:

$$\text{ENC}_{\mathbf{x}_i \leftarrow m_i} : \{1, 2, \dots, |\mathcal{M}_i|\} \longrightarrow \mathcal{X}_i^N \tag{13}$$

$$\text{DEC}_{\hat{m}_i \leftarrow \mathbf{y}_i} : \mathcal{Y}_i^N \longrightarrow \{0, 1, 2, \dots, |\mathcal{M}_i|\} \ , \tag{14}$$

for $i = 1, 2$. Again, we allow for the output of the channel decoding process to be 0 to indicated a (detected) error. Here the rates

$$R_i = \frac{\ln |\mathcal{M}_i|}{N} \ , \quad i = 1, 2 \ , \tag{15}$$

both in nats per parallel channel use, are key parameters of the system.

The source encoder consists of two mappings

$$\text{ENC}_{m_i \leftarrow \mathbf{s}} : \mathcal{S}^K \longrightarrow \{1, 2, \dots, |\mathcal{M}_i|\} \ , \quad i = 1, 2 \ . \tag{16}$$

The source decoder can be viewed as four separate mappings, depending upon whether or not there are channel decoding errors on the individual channels. Specifically, the source decoder can be constructed

$\hat{m}_2$

| $\hat{m}_1$ | $\hat{m}_2$ | $\mathrm{DEC}_{\hat{\mathbf{s}} \leftarrow \hat{m}}$ |
|:---:|:---:|:---:|
| $= 0$ | $= 0$ | $\mathrm{DEC}_{\hat{\mathbf{s}}_{\emptyset} \leftarrow \hat{m}_{\emptyset}}$ |
| $= 0$ | $\neq 0$ | $\mathrm{DEC}_{\hat{\mathbf{s}}_1 \leftarrow \hat{m}_1}$ |
| $\neq 0$ | $= 0$ | $\mathrm{DEC}_{\hat{\mathbf{s}}_2 \leftarrow \hat{m}_2}$ |
| $\neq 0$ | $\neq 0$ | $\mathrm{DEC}_{\hat{\mathbf{s}}_{1,2} \leftarrow \hat{m}_{1,2}}$ |

TABLE I

SOURCE CODING DIVERSITY DECODER RULES BASED UPON CHANNEL CONDITIONS.

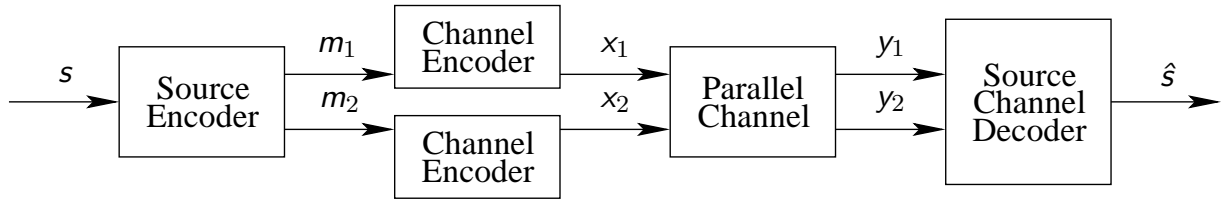

Fig. 6. Source coding diversity with joint source-channel decoding.

from the following four mappings:

$$\mathrm{DEC}_{\hat{\mathbf{s}}_{\emptyset} \leftarrow \hat{m}_{\emptyset}} : \{0\} \times \{0\} \longrightarrow \{s_*\}^K \tag{17}$$

$$\mathrm{DEC}_{\hat{\mathbf{s}}_1 \leftarrow \hat{m}_1} : \{1, 2, \ldots, |\mathcal{M}_1|\} \times \{0\} \longrightarrow \hat{\mathcal{S}}_1^K \tag{18}$$

$$\mathrm{DEC}_{\hat{\mathbf{s}}_2 \leftarrow \hat{m}_2} : \{0\} \times \{1, 2, \ldots, |\mathcal{M}_2|\} \longrightarrow \hat{\mathcal{S}}_2^K \tag{19}$$

$$\mathrm{DEC}_{\hat{\mathbf{s}}_{1,2} \leftarrow \hat{m}_{1,2}} : \{1, 2, \ldots, |\mathcal{M}_1|\} \times \{1, 2, \ldots, |\mathcal{M}_2|\} \longrightarrow \hat{\mathcal{S}}_0^K , \tag{20}$$

where $s_*$ is a constant determined by the distortion measure for the source; for example, if mean-square distortion is important, then $s_* = \mathrm{E}\,[s]$. Tab. I summarizes how these mappings are employed.

*4) Source Coding Diversity with Joint Decoding:* Finally, we also consider source coding diversity with joint decoding, as depicted in Fig. 6. Here all is the same as in the source coding diversity model of Fig. 5, except that source and channel decoding is performed jointly across channels by accounting for correlation among the channel coding inputs $m_1$ and $m_2$. Specifically, the channel decoding for this approach is a mapping

$$\mathrm{DEC}_{\hat{m}_{1,2} \leftarrow \mathbf{y}_{1,2}} : \mathcal{Y}_1^N \times \mathcal{Y}_2^N \longrightarrow \{0, 1, 2, \ldots, |\mathcal{M}_1|\} \times \{0, 1, 2, \ldots, |\mathcal{M}_1|\} \tag{21}$$

which also takes into account knowledge of the source coding structure. In practice full joint-design of the decoder may not be required and a partially separated design where likelihood-ratios, quantized likelihood-ratios or similar information are exchanged between the source and channel decoders may be sufficient.

### E. High-Resolution Approximations for Source Coding

An important practical example of our source model is the Gaussian source, for which $p_s(s)$ is a Gaussian density function with zero mean and unit variance. The Gaussian source also serves as a useful approximation to other sources in the high resolution (low distortion) regime [23], [50]. We now summarize the well-known results for single- and multiple-description source coding for the Gaussian case, and generalize them using the high resolution distortion approximations. These high resolution approximations are utilized throughout the sequel in our performance analysis.

*1) Single Description Source Coding:* In SD source coding, or classical rate-distortion theory, the source, $\mathbf{s}$, is quantized into a single description, $\hat{\mathbf{s}}$, using rate $R$.

In general, the rate-distortion function is difficult to determine, but a number of researchers have determined the rate-distortion function in the high resolution limit. Specifically, under some mild technical conditions [50],

$$\lim_{D \to 0} R(D) - \frac{1}{2} \log \frac{e^{2h(s)}}{2\pi e D} = 0 \ . \tag{22}$$

This result also implies that[3]

$$R(D) \approx \frac{1}{2} \log \frac{e^{2h(s)}}{2\pi e D} \tag{23}$$

Without loss of generality we scale a given source under consideration so that $e^{2h(s)} = 2\pi e$ to simplify the notation. Furthermore, instead of measuring the quantization rate in bits, we will find it more convenient to measure the rate in nats per channel sample by using the processing gain $\beta$ defined in Section II-D.1. Thus we will use the expressions

$$R(D) \approx \frac{1}{2\beta} \ln \frac{1}{D} \quad \text{and} \quad \exp R(D) \approx D^{-1/(2\beta)} \tag{24}$$

to approximate $R(D)$ and $\exp R(D)$ in high-resolution.

---

[3]Throughout the paper, the approximation $f(x) \approx g(x)$ is in the sense that $f(x)/g(x) \to 1$ and $|f(x) - g(x)| \to 0$ as $x$ approaches a limit, either $x \to 0$ or $x \to \infty$, which should be clear from the context.

As is well-known, the rate (in nats/channel sample) required for SD source coding of a Gaussian source at average distortion $D$ for any resolution is [36]

$$R_{\mathrm{sd}}(D) = \frac{1}{2\beta} \log \frac{1}{D} \ . \tag{25}$$

Therefore, one way to interpret (23), is that for difference distortion measures in the high-resolution limit all sources essentially look Gaussian except for scaling by the constant factor $\exp[2h(s)]/(2\pi e)$. Note that the form of the rate-distortion function in (23) is asymptotically accurate and not a worst case result like those in [51], [52].

*2) Multiple Description Source Coding:* In contrast to SD coding, MD source coding quantizes the source into two descriptions, $\hat{s}_1$ and $\hat{s}_2$ so that if only one is received then moderate distortion is incurred, and if both descriptions are received then lower distortion is obtained [28].

The rates and distortions achievable by coding a unit variance Gaussian source into two equal-rate descriptions with a total rate of $R_{\mathrm{md}}$ nats per channel sample, (*i.e.*, each description requires $R_{\mathrm{md}}/2$ nats) satisfy [28]

$$R_{\mathrm{md}}(D_0, D_1) = \frac{1}{2\beta} \log \frac{1}{D_0} + \frac{1}{2\beta} \log \frac{(1-D_0)^2}{(1-D_0)^2 - (1-2D_1+D_0)^2} \ , \tag{26a}$$

in the case of low distortions ($2D_1 - D_0 \leq 1$) where $D_0$ is the distortion when both descriptions are received and $D_1$ is the description when only a single description is received. For high distortions with ($2D_1 - D_0 \geq 1$), there is no penalty for the multiple descriptions and the total rate required is

$$R_{\mathrm{md}}(D_0, D_1) = \frac{1}{2\beta} \log \frac{1}{D_0}. \tag{26b}$$

The general rate-distortion region for the MD coding problem is still unknown, in the Gaussian case for more than two descriptions, and for more general sources. In the high resolution limit the rate-distortion region is the same as for a Gaussian source with variance $\exp[2h(s)]/(2\pi e)$ [23]. Hence for our asymptotic analysis we use the rate distortion function in (26) for both Gaussian and non-Gaussian sources with $\exp[2h(s)]/(2\pi e) = 1$.

Exponentiating (26a) yields

$$\exp[R_{\mathrm{md}}(D_0, D_1)] = D_0^{-1/(2\beta)} \cdot (1-D_0)^{-1/\beta}$$
$$\cdot (1 - 2D_0 + D_0^2 - 1 - D_0^2 - 2D_0 + 4D_1 + 4D_0 D_1 - 4D_1^2)^{-1/(2\beta)} \tag{27}$$

$$= D_0^{-1/(2\beta)} \cdot (1-D_0)^{-1/(\beta)} \cdot (4D_1 - 4D_0 + 4D_0 D_1 - 4D_1^2)^{-1/(2\beta)} \tag{28}$$

$$\approx D_0^{-1/(2\beta)} \cdot (4D_1 - 4D_0)^{-1/(2\beta)} \tag{29}$$

where the last line follows since $(1 - D_0) \approx 1$ and $4(D_1 - D_0 + D_0 D_1 - D_1^2) \approx 4(D_1 - D_0)$ as $D_0 \to 0$ and $D_1 \to 0$. If only $D_0 \to 0$, then the $\approx$ in (29) must be replaced with $\gtrapprox$. Any reasonable multiple description system has $D_0 \leq D_1/2$ (otherwise the denominator of (26a) could be easily increased while decreasing the distortion by setting $D_1 = 2D_0$). So since $2D_1 \leq 4(D_1 - D_0) \leq 4D_1$ we obtain

$$(4D_0 D_1)^{-1/(2\beta)} \lessapprox \exp[R_{\mathrm{md}}(D_0, D_1)] \lessapprox (2D_0 D_1)^{-1/(2\beta)} \tag{30}$$

where the lower bound holds when $D_0 \to 0$ and the upper bound also requires $D_1 \to 0$.

## III. On-Off Component Channels

In this section, we examine the performance of source and channel coding diversity for scenarios in which each of the component channels is either "on", supporting a given transmission rate, or "off", supporting no rate (or an arbitrarily small rate). Much of the literature suggests that source coding diversity was developed for, and performs well on, such channel models. Our analysis is based upon channels that are parameterized in a manner similar to the continuous channels in Section IV. This parameterization allows us to compare source and channel coding diversity over a broad range of operating conditions. In addition to confirming that there exist operating conditions for which source coding diversity significantly outperforms channel coding diversity, our results illustrate that there also exist operating conditions for which the performance difference between source and channel coding diversity is negligible.

### A. Component Channel Model

For cases in which we are concerned with prolonged, deep fading or shadowing in a mobile radio channel, strong first-adjacent interference in a terrestrial broadcast channel, or congestion in a network, we can model the channel state $a_i$ as taking on only two possible values. Specifically, we can consider on-off channels where the channel mutual information has probability law

$$I = \begin{cases} \ln(1 + \mathrm{SNR}) \ , & \text{with probability } (1 - \epsilon) \\ 0 \ , & \text{with probability } \epsilon \end{cases} . \tag{31}$$

In (31), SNR parameterizes the channel quality when the channel is on, and $\epsilon$ parameterizes the probability that the channel is off. There is no connection between the channels' probability of being off and the quality in the on state; that is, neither SNR nor the selected encoding rate $R$ effects $\epsilon$. By contrast, for the continuous channels discussed in Section IV, $\epsilon$ will depend directly on both.

For simplicity of exposition, and ease of comparison with continuous channel scenarios in the sequel, the term *outage* will refer to the inability of a given approach to convey information over the pair of component channels. If both channels are off, then the system experiences outage regardless of the communication approach; however, as we will see, different approaches may or may not experience outage when one of the channels is on and the other is off. For all of the approaches we discuss, due to the nature of the on-off channels, performance can be classified into two regimes. The *quality-limited regime* has average distortion performance varying dramatically with the channel quality in the on state, because the distortion under no outage dominates the average distortion. In this case, the distortion under no outage is limited by the rate communicated, which, in turn, is limited by the channel quality. The *outage-limited regime* has average distortion performance that does not vary dramatically with the channel quality in the on state, because the distortion under outage dominates the average distortion.

### B. No Diversity

Combining a SD source coder with a single component channel with channel encoder and decoder, the average distortion, as a function of the source coding rate $R$, is given by

$$\mathrm{E}\left[D_{\mathrm{NO-DIV}}(R)\right] = \begin{cases} (1-\epsilon)\exp(-2\beta) + \epsilon \ , & \text{if } 0 < R \leq \ln(1+\mathrm{SNR}) \\ 1 \ , & \text{otherwise} \end{cases} . \qquad (32)$$

Thus, the minimum average distortion is

$$D_{\mathrm{NO-DIV}} = \min_R \mathrm{E}\left[D_{\mathrm{NO-DIV}}(R)\right]$$

$$= (1-\epsilon)(1+\mathrm{SNR})^{-2\beta} + \epsilon \ . \qquad (33)$$

We say that this system operates in the quality-limited regime if

$$(1+\mathrm{SNR})^{2\beta} \ll \frac{1-\epsilon}{\epsilon} \ , \qquad (34)$$

in which case, the average distortion behaves essentially as $(1-\epsilon)(1+\mathrm{SNR})^{-2\beta}$. If

$$(1+\mathrm{SNR})^{2\beta} \gg \frac{1-\epsilon}{\epsilon} \ , \qquad (35)$$

the system operates in the outage-limited regime, in which case the average distortion behaves essentially as $\epsilon$.

### C. Optimal Channel Coding Diversity

Combining a SD source coder with optimal parallel channel coding over the component channels, the average distortion, as a function of the source coding rate $R$, is given by

$$
\mathrm{E}\left[D_{\mathrm{OPT-CCDIV}}(R)\right] = \begin{cases} (1-\epsilon^2)\exp(-2\beta R) + \epsilon^2\ , & \text{if } 0 < R \leq \ln(1+\mathrm{SNR}/2) \\ (1-\epsilon)^2\exp(-2\beta R) + [1-(1-\epsilon)^2]\ , & \text{if } \ln(1+\mathrm{SNR}/2) < R \leq 2\ln(1+\mathrm{SNR}/2) \\ 1\ , & \text{otherwise} \end{cases} \cdot
$$

(36)

For parallel channel coding, the two channel codewords are independent, and the system is able to sum the mutual informations of the component channels. This leads to the upper bound of $R \leq 2\ln(1+\mathrm{SNR}/2)$ in the second case of (36). If we instead utilized repetition coding, so that the two channel codewords are identical, the upper bound in the second case would instead be $R \leq \ln(1+\mathrm{SNR})$.

In contrast to the case of no diversity, the performance of the optimal channel coding diversity exhibits a discontinuity as a function of $R$. Fig. 7 illustrates that, because of the discrete probability distribution on the channel states, a discontinuity arises in the outage probability about the point $R = \ln(1+\mathrm{SNR}/2)$.

Clearly, each case in (36) is minimized by utilizing the largest possible rate for that case. Then the minimum average distortion becomes

$$
\begin{aligned}
D_{\mathrm{OPT-CCDIV}} &= \min_R \mathrm{E}\left[D_{\mathrm{OPT-CCDIV}}(R)\right] \\
&= \min\Big\{ (1-\epsilon^2)(1+\mathrm{SNR}/2)^{-2\beta} + \epsilon^2\ , \\
&\qquad\quad (1-\epsilon)^2(1+\mathrm{SNR}/2)^{-4\beta} + [1-(1-\epsilon)^2] \Big\}\ .
\end{aligned}
$$

(37)

As Fig. 8 illustrates, the two terms in (37) have their own quality- and outage-limited regimes, which, when combined by the minimum operation, leads to four trends in the overall system performance.

Comparing the two terms in (37), we see that the different choices of rate lead to different costs and benefits. Using the lower transmission rate, $R = \ln(1+\mathrm{SNR}/2)$, (*cf.* the first term in (37)) results in better outage-limited performance, but worse quality-limited performance. This approach exploits the diversity gain of the underlying parallel channel. On the other hand, using the higher transmission rate, $R = 2\ln(1+\mathrm{SNR}/2)$, (*cf.* the second term in (37)) results in worse outage-limited performance, but better quality-limited performance. This approach exploits the multiplexing gain of the underlying parallel channel. We note that the diversity and multiplexing terminology is inspired by the inherent tradeoff between the two for multiple-input, multiple-output (MIMO) wireless systems operating over fading channels [48].

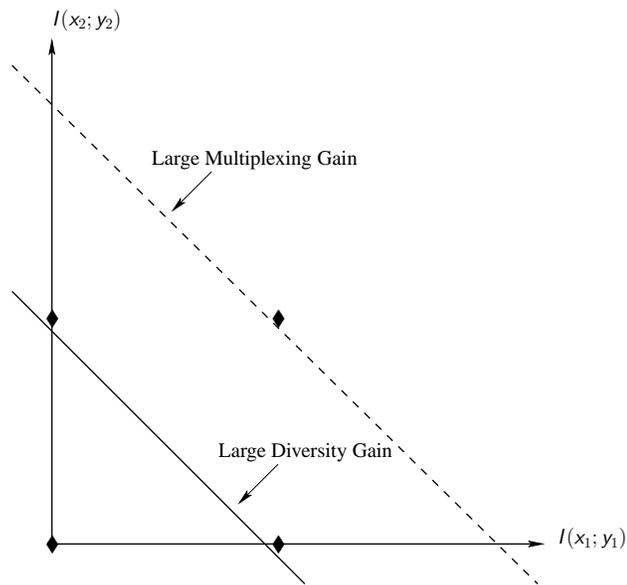

Fig. 7. Outage region boundaries for optimal parallel channel coding. The ◆ symbols correspond to the sample mutual information pairs $(0,0)$, $(0, \ln(1 + \text{SNR}/2))$, $(\ln(1 + \text{SNR}/2), 0)$, and $(\ln(1 + \text{SNR}/2), \ln(1 + \text{SNR}/2))$. The solid line corresponds to the first case of (36), in which a low rate is selected to take advantage of diversity gain. The dashed line corresponds to the second case of (36), in which a higher rate is selected to take advantage of multiplexing gain. Outage regions are below and to the left of these diagonals.

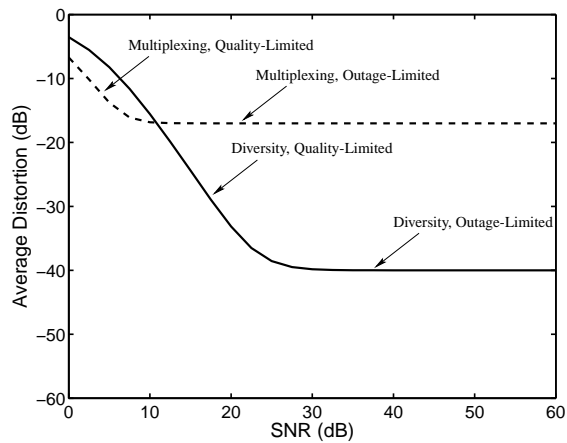

Fig. 8. Average distortion performance with $\epsilon = 10^{-2}$ for the first (solid line) and second (dashed line) terms in the minimization of (37).

Note that the two terms in (37) are equal when

$$(1 + \text{SNR}/2)^{2\beta} = \frac{1 - \epsilon}{2\epsilon} \ . \tag{38}$$

For small SNR (such that $(1+\text{SNR}/2)^{2\beta} < (1-\epsilon)/(2\epsilon)$), we exploit the multiplexing mode of operation and pass through its quality-limited and outage-limited regimes as we increase SNR until (38) is satisfied. As we will see, passing through the outage-limited regime of the multiplexing mode is the key limitation of optimal channel coding diversity for on-off channels. For higher SNR (such that $(1 + \text{SNR}/2)^{2\beta} > (1-\epsilon)/(2\epsilon)$), we exploit the diversity mode of operation and pass through its quality- and outage-limited regimes as we increase SNR.

### D. Source Coding Diversity

In this section, we approximate the minimum average distortion for an MD system with independent channel coding. The analysis of this system is slightly more involved than those of previous sections because the rate-distortion region for MD coding is more complex, and independent channel coding over on-off component channels involves a pair of outage events.

Similar to Fig. 7, Fig. 9 displays outage region boundaries for independent channel coding. It is straightforward to see that the source coder should employ rates no greater than $\ln(1+\text{SNR}/2)$ on each of the component channels; otherwise, one of the channels exhibits outage with probability one, and the system can perform no better than the case of no diversity with half the SNR. As a result, our analysis only considers the case $R_i \leq \ln(1+\text{SNR}/2)$. Moreover, due to the symmetry of the component channels, one can expect symmetric rates, $i.e.$, $R_1 = R_2 = R$, to be optimal; thus, we focus on this case. With these simplifications, we observe that, in contrast to the triangular outage regions for optimal parallel channel coding in Fig. 7, the rectangular outage regions for independent channel coding in Fig. 9 are well-matched to the on-off channel realizations.

Optimizing average distortion for the MD system requires a tradeoff between the distortion $D_1 = D_2$ achieved when only one description is received and the joint distortion $D_0$ achieved when both descriptions are received. Although this tradeoff is available in (30), we refactor it for our purposes here. Specifically, we set

$$D_1 = D_2 \approx \begin{cases} \exp(-(1-\lambda)2\beta R) \ , & 0 \leq \lambda < 1 \\ 1 \ , & \lambda = 1 \end{cases} \ , \tag{39}$$

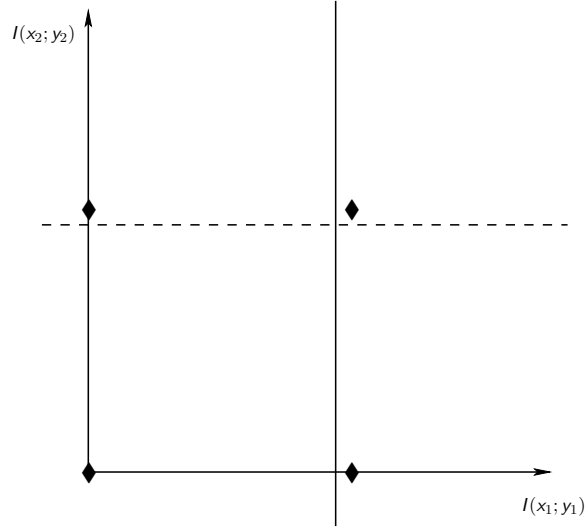

Fig. 9.   Outage region boundaries for MD source coding with independent channel coding. The ◆ symbols correspond to the sample mutual information pairs $(0, 0)$, $(0, \ln(1 + \text{SNR}/2))$, $(\ln(1 + \text{SNR}/2), 0)$, and $(\ln(1 + \text{SNR}/2), \ln(1 + \text{SNR}/2))$. The solid line corresponds to the outage region boundary for the first channel, and the dashed line corresponds to the outage region boundary for the second channel. The outage region for channel one (resp. channel two) is to the left (resp. below) the boundary.

where $R$ is the channel coding rate for a single channel. Thus, if $\lambda = 0$, the individual descriptions achieve the single description rate-distortion bound. With this parameterization of $D_1$ and $D_2$, the MD high-resolution approximation (30) yields

$$D_0 \approx \begin{cases} \frac{1}{2} \exp(-(1 + \lambda)2\beta R) \ , & 0 \leq \lambda < 1 \\ \exp(-4\beta R) \ , & \lambda = 1 \end{cases} \tag{40}$$

for the joint distortion when both descriptions are received. We note that an essentially identical approximation is developed in [16].

The minimum average distortion for source coding diversity is then approximately

$$D_{\text{SCDIV}} \approx \min\{ \min_{0 < \lambda < 1} \epsilon^2 + 2\epsilon(1 - \epsilon)(1 + \text{SNR}/2)^{-(1-\lambda)2\beta} + \frac{1}{2}(1 - \epsilon)^2(1 + \text{SNR}/2)^{-(1+\lambda)2\beta},$$

$$[1 - (1 - \epsilon)^2] + (1 - \epsilon)^2(1 + \text{SNR}/2)^{-4\beta} \} \ . \tag{41}$$

For $\lambda = 1$, source coding diversity performance reduces to that of channel coding diversity; for $\lambda = 0$, source coding diversity performance reduces to that of no diversity with half the SNR. Because optimization over $\lambda$ does not lend much insight, we delay discussion of source coding diversity quality-and outage-limited regimes to the next section, where we also compare with the other approaches.

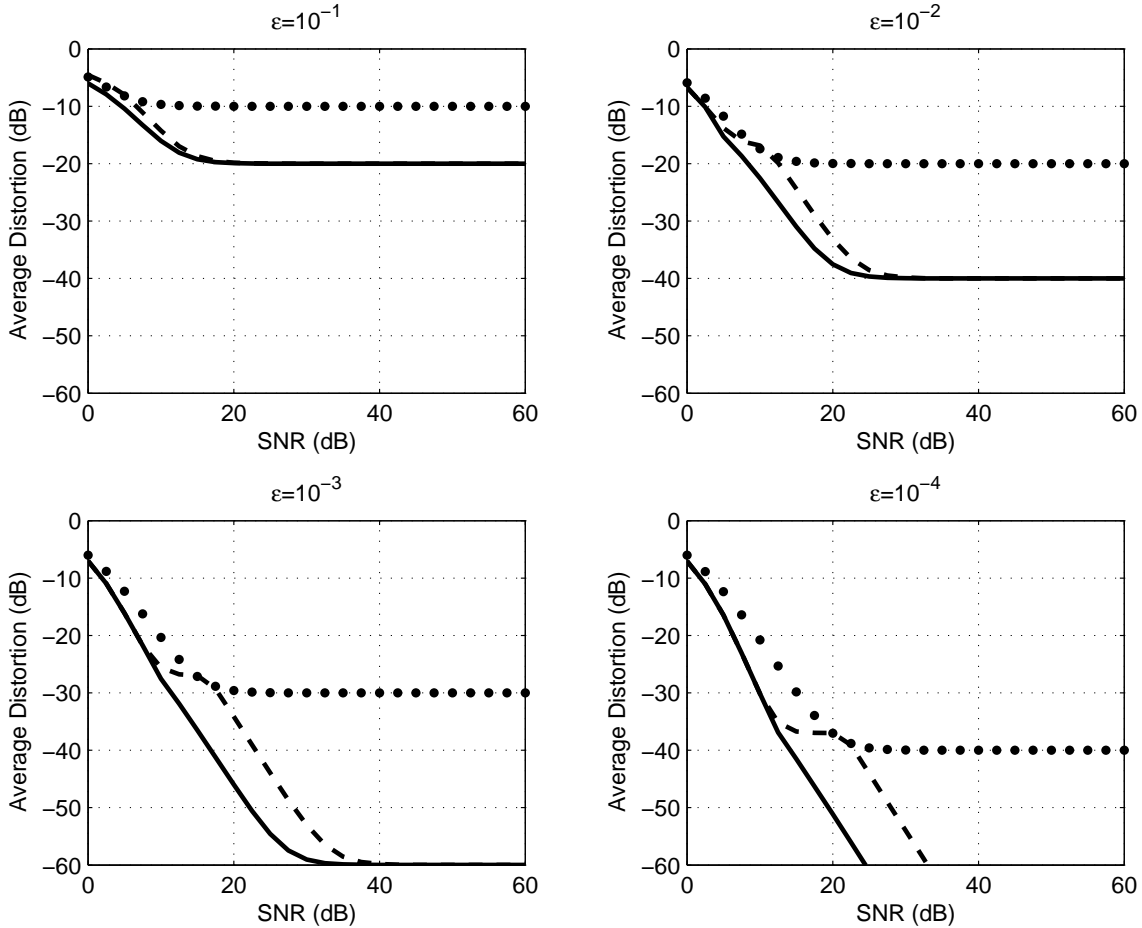

Fig. 10. Average distortion performance over on-off channels. The plots show average distortion as a function of SNR; successively lower curves correspond to no diversity (dotted lines), optimal channel coding diversity (dashed lines), and source coding diversity (solid lines), respectively. Each plot corresponds to a different value for the probability $\epsilon$ of a component channel being off, and all are for $\beta = 1$.

### E. Comparison

Fig. 10 compares average distortion performance of source and channel coding diversity by displaying the minimum average distortions (33), (37), and (41) as functions of the component channel quality, SNR, in the on state, for different values of the probability of a component channel being off, $\epsilon$. The results in Fig. 10 are clearly consistent with our intuitive discussion of source and channel coding diversity performance in Section I-A. For moderate SNR, depending upon $\epsilon$, both systems exhibit transitions from $\mathrm{SNR}^{-4}$ behavior to $\mathrm{SNR}^{-2}$ behavior; however, the transition is generally less drastic for source coding diversity, especially for smaller $\epsilon$. The difference between the two systems is apparently the outage-limited behavior of the multiplexing mode for optimal channel coding diversity, for which the outage regions are not well-matched to the channel realizations. By contrast, the transition between the two quality-limited trends for source coding diversity is much less drastic, and this graceful degradation property of source coding diversity leads to their better performance over on-off channels. However, it is important to note that there is negligible difference between optimal channel coding diversity and source coding diversity at both low and high SNR.

## IV. Continuous State Channels

In cases where we are concerned with time or frequency selective multipath fading in a mobile radio channel or a range of possible interference levels in a cellular network, we can model the channel state $a_i$ as taking on a continuum of values. For example, multiplicative fading is commonly modeled as a Rayleigh or Nakagami random variable in such scenarios. In the following section we study the average mean square distortion in the limit of high SNR for such continuous channels when the channel state is known to the receiver but not the transmitter. Since the distortion generally behaves as $\mathrm{SNR}^{-\Delta}$ for such channels, we are mainly interested in computing the distortion exponent defined as

$$\Delta = -\lim_{\mathrm{SNR}\to\infty} \frac{\log E[D]}{\log \mathrm{SNR}}. \tag{42}$$

Note that there is an important difference between the average or transmit signal-to-noise ratio which is deterministic and known by both transmitter and receiver and the instantaneous or block signal-to-noise ratio which is random and known only at the receiver. Throughout the rest of the paper, we always use SNR to refer to the former and consider the random, instantaneous signal-to-noise ratio as a random variable.

In Section IV-G, we plot the distortion exponents as well as the numerically computed average distortions for a Gaussian source transmitted over a complex Rayleigh fading additive white Gaussian noise channel. Hence the reader may find it useful to refer to Figures 11 and 12 as a concrete example for comparing the following results for the performance of each system.

### A. Continuous Channel Model

For continuous state channels, the distribution of the mutual information random variable is generally difficult to compute exactly. For complex, additive white Gaussian noise channels with multiplicative fading, however, the mutual information random variable is $I = \log(1 + a \cdot \text{SNR})$ where $a$ corresponds to the multiplicative fading which is normalized so that $E[a] = 1$ so that SNR is the transmit power or equivalently, the average received power. For $a \cdot \text{SNR} \gg 1$, we have

$$I = \log(a \cdot \text{SNR}) + \log\left(1 + \frac{1}{a \cdot \text{SNR}}\right) = \log(a \cdot \text{SNR}) + O\left(\frac{1}{a \cdot \text{SNR}}\right) \approx \log(a \cdot \text{SNR})$$

and so $\exp I$ is close to $\text{SNR} \cdot a$. [4] Thus, for additive Gaussian noise channels with multiplicative fading, we can develop asymptotic results by considering the first terms in the Taylor series expansion of the distribution of $a$ near zero. More generally, we can focus on the high SNR limit by considering the Taylor series expansion of the distribution for the mutual information random variable for each channel.

Specifically, let $f_I(t)$ and $F_I(t)$ represent the probability density function (PDF) and cumulative distribution function (CDF) for the mutual information and let $f_{e'}(t)$ and $F_{e'}(t)$ represent the PDF and CDF for $I$.[5] We consider the case where there exists a parameter called SNR such that

$$f_{e'}(t) \approx c p \left(\frac{t}{\text{SNR}}\right)^{p-1} \quad \text{(with } p \geq 1\text{)} \tag{43}$$

and consequently $F_{e'}(t)$ can be approximated via

$$F_{e'}(t) \approx c \left(\frac{t}{\text{SNR}}\right)^{p}. \tag{44}$$

Intuitively, SNR represents the transmit signal-to-noise ratio or the average signal-to-noise ratio and $F_{e'}(t)$ is the probability that the instantaneous signal-to-noise ratio is below $t$. As introduced in Section II-E.1, the notion of approximation we use is that $a(\text{SNR}) \approx b(\text{SNR})$ if $\lim_{\text{SNR} \to \infty} a(\text{SNR})/b(\text{SNR}) = 1$ and $\lim_{\text{SNR} \to \infty} |a(\text{SNR}) - b(\text{SNR})| = 0$.

For example, in wireless communications, a common model is an additive white Gaussian noise channel with fading:

$$y[i] = a \cdot x[i] + z[i] \tag{45}$$

where $a$ represents the fading and $z[i]$ represents additive noise. A common approach is to obtain robustness by coding over two separate frequency bands or time-slots in which case the channel model

---

[4]A similar expression can also be obtained for additive noise channels with non-Gaussian noise (*e.g.*, using techniques from [53], [54]).

[5]Recall that we assume the mutual information optimizing input distribution is independent of the channel state. Hence it makes sense to speak of the mutual information distribution as given instead of a parameter controlled by the system designer.

becomes

$$y_1[i] = a_1 \cdot x_1[i] + z_1[i]$$

$$y_2[i] = a_2 \cdot x_2[i] + z_2[i].$$

If we are interested in Rayleigh fading then each $a_i$ has an exponential distribution and at high SNR, the cumulative distribution function for $\exp I(\mathbf{y}_i; \mathbf{x}_i)$ is approximated by $t/\text{SNR}$ and hence the parameters $c$ and $p$ in (44) are both unity (*e.g.*, see [55], [56] for a discussion of such high SNR expansions).

## B. No Diversity

Perhaps the simplest case to consider is when there is only a single channel and no diversity is present. For such a scenario, a natural approach is cascading an SD source encoder/decoder $\text{ENC}_{m \leftarrow \mathbf{s}}(\cdot)/\text{DEC}_{\hat{\mathbf{s}} \leftarrow \hat{m}}(\cdot)$ with a single channel encoder/decoder $\text{ENC}_{\mathbf{x} \leftarrow m}(\cdot)/\text{DEC}_{\hat{m} \leftarrow \mathbf{y}}(\cdot)$. In terms of our general joint source-channel coding notation such a system has the encoder and decoder

$$\mathbf{x} = \text{ENC}_{\mathbf{x} \leftarrow \mathbf{s}}(\mathbf{s}) = \text{ENC}_{\mathbf{x} \leftarrow m}(\text{ENC}_{m \leftarrow \mathbf{s}}(\mathbf{s})) \tag{46a}$$

$$\hat{\mathbf{s}} = \text{DEC}_{\hat{\mathbf{s}} \leftarrow \mathbf{y}}(\mathbf{y}) = \begin{cases} \text{DEC}_{\hat{\mathbf{s}} \leftarrow \hat{m}}(\text{DEC}_{\hat{m} \leftarrow \mathbf{y}}(\mathbf{y})), & \text{DEC}_{\hat{m} \leftarrow \mathbf{y}}(\mathbf{y}) \neq 0 \\ E[\mathbf{s}], & \text{otherwise.} \end{cases} \tag{46b}$$

*Theorem 1:* The distortion exponent for a system with no diversity described by (46) is

$$\Delta_{\text{NO-DIV}} = \frac{2\beta p}{2\beta + p} , \tag{47}$$

where $\beta$ is the processing gain defined in Section II-D.1 and $p$ is the diversity order of the channel approximation in (44).

*Proof:* The average distortion is

$$E[D] = \min_D \Pr[I(x; y) < R(D)] + \{1 - \Pr[I(x; y) < R(D)]\} \cdot D \tag{48}$$

$$= \min_D F_{e'}(\exp R(D)) + [1 - F_{e'}(R(D))] \cdot D \tag{49}$$

$$\approx \min_D c\frac{D^{-p/(2\beta)}}{\text{SNR}^p} + \left[1 - c\frac{D^{-p/(2\beta)}}{\text{SNR}^p}\right] \cdot D \tag{50}$$

$$\approx \min_D c\frac{D^{-p/(2\beta)}}{\text{SNR}^p} + D. \tag{51}$$

Differentiating and setting equal to 0 yields the minimizing distortion

$$D^* = \left(\frac{2\beta}{cp}\right)^{\frac{-2\beta}{2\beta + p}} \cdot \text{SNR}^{\frac{-2\beta p}{2\beta + p}}.$$

Substituting this into (51) yields

$$E[D] \approx C_{\text{NO}-\text{DIV}} \cdot \text{SNR}^{\frac{-2\beta p}{2\beta + p}}. \tag{52}$$

where $C_{\text{NO}-\text{DIV}}$ represents a term independent of SNR. Thus the distortion exponent is $2\beta p/(2\beta + p)$.

■

### C. Selection Channel Coding Diversity

Perhaps the simplest approach to using two independent channels is to use SD source coding with repetition channel coding and selection combining. In this scheme, the encoder quantizes the source, $\mathbf{s}$, to $\hat{\mathbf{s}}$, adds channel coding to produce $\mathbf{x}$, and repeats the result on both channels. The receiver decodes the higher quality channel and ignores the other. Formally, the encoder and decoder are given by

$$(\mathbf{x}_1, \mathbf{x}_2) = \text{ENC}_{\mathbf{x}_1, \mathbf{x}_2 \leftarrow \mathbf{s}}(\mathbf{s}) = (\text{ENC}_{\mathbf{x} \leftarrow m}(\text{ENC}_{m \leftarrow \mathbf{s}}(\mathbf{s})), \text{ENC}_{\mathbf{x} \leftarrow m}(\text{ENC}_{m \leftarrow \mathbf{s}}(\mathbf{s}))) \tag{53a}$$

$$\hat{\mathbf{s}} = \text{DEC}_{\hat{\mathbf{s}} \leftarrow \mathbf{y}_1, \mathbf{y}_2}(\mathbf{y}_1, \mathbf{y}_2) = \begin{cases} \text{DEC}_{\hat{\mathbf{s}} \leftarrow \hat{m}}(\text{DEC}_{\hat{m} \leftarrow \mathbf{y}}(\mathbf{y}_1)), & \text{DEC}_{\hat{m} \leftarrow \mathbf{y}}(\mathbf{y}_1) \neq 0 \\ \text{DEC}_{\hat{\mathbf{s}} \leftarrow \hat{m}}(\text{DEC}_{\hat{m} \leftarrow \mathbf{y}}(\mathbf{y}_2)), & \text{DEC}_{\hat{m} \leftarrow \mathbf{y}}(\mathbf{y}_1) = 0 \text{ and } \text{DEC}_{\hat{m} \leftarrow \mathbf{y}}(\mathbf{y}_2) \neq 0 \\ E[\mathbf{s}], & \text{otherwise} \end{cases}$$

$$\tag{53b}$$

where $\text{ENC}_{m \leftarrow \mathbf{s}}(\cdot)/\text{DEC}_{\hat{\mathbf{s}} \leftarrow \hat{m}}(\cdot)$ correspond to the SD source encoder/decoder and $\text{ENC}_{\mathbf{x} \leftarrow m}(\cdot)/\text{DEC}_{\hat{m} \leftarrow \mathbf{y}}(\cdot)$ correspond to the single channel encoder/decoder. Thus, the quantized source signal will be recovered provided either channel is good. While such a scheme is sub-optimal in terms of resource use, it is simplest to understand and easiest to implement. The following theorem (proved in Appendix A) characterize asymptotic performance.

*Theorem 2:* The distortion exponent for a system with selection channel coding diversity described by (53) is

$$\Delta_{\text{SEL}-\text{CCDIV}} = \frac{2\beta p}{\beta + p}. \tag{54}$$

### D. Multiplexed Channel Coding Diversity

A key drawback of repetition coding with selection combining is that it wastes the potential bandwidth of one channel in order to provide diversity. When the channel is usually good, such a scheme can be significantly sub-optimal. Hence, a complementary approach is channel multiplexing where the source is quantized using SD coding and this message is split over both channels. We define a channel multiplexing system as one with encoder and decoder given by

$$(\mathbf{x}_1, \mathbf{x}_2) = \mathrm{ENC}_{\mathbf{x}_1, \mathbf{x}_2 \leftarrow \mathbf{s}}(\mathbf{s}) = (\mathrm{ENC}_{\mathbf{x}_1 \leftarrow m_1}(\mathrm{ENC}_{m_1 \leftarrow \mathbf{s}}(\mathbf{s})), \mathrm{ENC}_{\mathbf{x}_2 \leftarrow m_2}(\mathrm{ENC}_{m_2 \leftarrow \mathbf{s}}(\mathbf{s}))) \qquad (55a)$$

$$\hat{\mathbf{s}} = \mathrm{DEC}_{\hat{\mathbf{s}} \leftarrow \mathbf{y}_1, \mathbf{y}_2}(\mathbf{y}_1, \mathbf{y}_2) = \begin{cases} \mathrm{DEC}_{\hat{\mathbf{s}} \leftarrow \hat{m}}(\mathrm{DEC}_{\hat{m}_1 \leftarrow \mathbf{y}_1}(\mathbf{y}_1), \mathrm{DEC}_{\hat{m}_2 \leftarrow \mathbf{y}_2}(\mathbf{y}_2)), & \mathrm{DEC}_{\hat{m}_1 \leftarrow \mathbf{y}_1}(\mathbf{y}_1) \neq 0 \text{ and} \\ & \mathrm{DEC}_{\hat{m}_2 \leftarrow \mathbf{y}_2}(\mathbf{y}_2) \neq 0 \\ E[\mathbf{s}], & \text{otherwise.} \end{cases}$$

$$(55b)$$

where $\mathrm{ENC}_{\mathbf{x}_i \leftarrow m_i}(\cdot)/\mathrm{DEC}_{\hat{m}_i \leftarrow \mathbf{y}_i}(\cdot)$ correspond to single channel encoders/decoders and $\mathrm{ENC}_{m_i \leftarrow \mathbf{s}}(\cdot)$ correspond to the first and second half of the output of a single description source encoder with decoder $\mathrm{DEC}_{\hat{\mathbf{s}} \leftarrow \hat{m}}(\cdot)$. If both channels are good enough to support successful decoding, then this scheme can transmit roughly twice the rate of a repetition coding system. The drawback is since either channel being bad can cause decoding failure, the system is less robust. The following theorem (proved in Appendix B) characterizes asymptotic performance.

*Theorem 3:* The distortion exponent for a system with multiplexed channel coding diversity described by (55) is

$$\Delta_{\mathrm{MPX-CCDIV}} = 4p\beta/(p + 4\beta). \qquad (56)$$

Intuitively, we expect that when bandwidth is plentiful and outage is the dominating concern, the diversity provided by repetition coding is more important than the extra rate provided by channel multiplexing. When bandwidth is scarce, we expect the reverse to be true. We can verify this intuition by examining the distortion exponents in these two limits to obtain

$$\lim_{\beta/p \to \infty} \frac{\Delta_{\mathrm{SEL-CCDIV}}}{\Delta_{\mathrm{MPX-CCDIV}}} = 2 \qquad (57)$$

$$\lim_{\beta/p \to 0} \frac{\Delta_{\mathrm{SEL-CCDIV}}}{\Delta_{\mathrm{MPX-CCDIV}}} = \frac{1}{2}. \qquad (58)$$

The distortion exponents are equal if $p = 2\beta$.

### E. Optimal Channel Coding Diversity

Each of the previous schemes used SD source coding with some form of independent channel coding and hence was sub-optimal. With SD source coding, the optimal strategy is to use parallel channel coding. In this scheme, the two component channels are treated as a single parallel channel with channel encoding and decoding performed jointly over both. Specifically, we define optimal channel coding diversity as

$$(\mathbf{x}_1, \mathbf{x}_2) = \text{ENC}_{\mathbf{x}_1, \mathbf{x}_2 \leftarrow \mathbf{s}}(\mathbf{s}) = \text{ENC}_{\mathbf{x} \leftarrow m}(\text{ENC}_{m \leftarrow \mathbf{s}}(\mathbf{s})) \tag{59a}$$

$$\hat{\mathbf{s}} = \text{DEC}_{\hat{\mathbf{s}} \leftarrow \mathbf{y}_1, \mathbf{y}_2}(\mathbf{y}_1, \mathbf{y}_2) = \begin{cases} \text{DEC}_{\hat{\mathbf{s}} \leftarrow \hat{m}}(\text{DEC}_{\hat{m} \leftarrow \mathbf{y}}(\mathbf{y}_1, \mathbf{y}_2)), & \text{DEC}_{\hat{m} \leftarrow \mathbf{y}}(\mathbf{y}_1, \mathbf{y}_2) \neq 0 \\ E[\mathbf{s}], & \text{otherwise} \end{cases} \tag{59b}$$

where $\text{ENC}_{m \leftarrow \mathbf{s}}(\cdot)/\text{DEC}_{\hat{\mathbf{s}} \leftarrow \hat{m}}(\cdot)$ correspond to the SD source encoder/decoder and $\text{ENC}_{\mathbf{x} \leftarrow m}(\cdot)/\text{DEC}_{\hat{m} \leftarrow \mathbf{y}}(\cdot)$ correspond to the parallel channel encoder/decoder. Since parallel channel coding optimally uses the channel resources, it dominates both repetition coding with selection combining and channel multiplexing as characterized by the following theorem (proved in Appendix C).

*Theorem 4:* The distortion exponent for a system with optimal channel coding diversity described by (59) is

$$\Delta_{\text{OPT−CCDIV}} = \frac{4p\beta}{p + 2\beta}. \tag{60}$$

### F. Source Coding Diversity

Next, we consider the case where the source is transmitted over a pair of independent channels using MD source coding. Specifically, we consider a system with

$$(\mathbf{x}_1, \mathbf{x}_2) = \text{ENC}_{\mathbf{x}_1, \mathbf{x}_2 \leftarrow \mathbf{s}}(\mathbf{s}) = (\text{ENC}_{\mathbf{x}_1 \leftarrow m_1}(\text{ENC}_{m_1 \leftarrow \mathbf{s}}(\mathbf{s})), \text{ENC}_{\mathbf{x}_2 \leftarrow m_2}(\text{ENC}_{m_2 \leftarrow \mathbf{s}}(\mathbf{s}))) \tag{61a}$$

$$\hat{\mathbf{s}} = \text{DEC}_{\hat{\mathbf{s}} \leftarrow \mathbf{y}_1, \mathbf{y}_2}(\mathbf{y}_1, \mathbf{y}_2) = \begin{cases} \text{DEC}_{\hat{\mathbf{s}}_1 \leftarrow \hat{m}_1}(\text{DEC}_{\hat{m}_1 \leftarrow \mathbf{y}_1}(\mathbf{y}_1)), & \text{DEC}_{\hat{m}_1 \leftarrow \mathbf{y}_1}(\mathbf{y}_1) \neq 0 \text{ and} \\ & \text{DEC}_{\hat{m}_2 \leftarrow \mathbf{y}_2}(\mathbf{y}_2) = 0 \\ \text{DEC}_{\hat{\mathbf{s}}_2 \leftarrow \hat{m}_2}(\text{DEC}_{\hat{m}_2 \leftarrow \mathbf{y}_2}(\mathbf{y}_2)), & \text{DEC}_{\hat{m}_1 \leftarrow \mathbf{y}_1}(\mathbf{y}_1) = 0 \text{ and} \\ & \text{DEC}_{\hat{m}_2 \leftarrow \mathbf{y}_2}(\mathbf{y}_2) \neq 0 \\ \text{DEC}_{\hat{\mathbf{s}}_{1,2} \leftarrow \hat{m}_{1,2}}(\text{DEC}_{\hat{m}_1 \leftarrow \mathbf{y}_1}(\mathbf{y}_1), \text{DEC}_{\hat{m}_2 \leftarrow \mathbf{y}_2}(\mathbf{y}_2)), & \text{DEC}_{\hat{m}_1 \leftarrow \mathbf{y}_1}(\mathbf{y}_1) \neq 0 \text{ and} \\ & \text{DEC}_{\hat{m}_2 \leftarrow \mathbf{y}_2}(\mathbf{y}_2) \neq 0 \\ E[\mathbf{s}], & \text{DEC}_{\hat{m}_1 \leftarrow \mathbf{y}_1}(\mathbf{y}_1) = 0 \text{ and} \\ & \text{DEC}_{\hat{m}_2 \leftarrow \mathbf{y}_2}(\mathbf{y}_2) = 0 \end{cases} \tag{61b}$$

where $\text{ENC}_{m_1 \leftarrow \mathbf{s}}(\cdot)$ and $\text{ENC}_{m_2 \leftarrow \mathbf{s}}(\cdot)$ represent the two quantizations of the source produced by the MD source coder, $\text{DEC}_{\hat{\mathbf{s}}_i \leftarrow \hat{m}_i}(\cdot)$ represent the possible source decoders described in Tab. I, and $\text{ENC}_{\mathbf{x}_i \leftarrow m_i}(\cdot)$ / $\text{DEC}_{\hat{m}_i \leftarrow \mathbf{y}_i}(\cdot)$ correspond to single channel encoders/decoders. The performance of such a system is characterized by Theorem 5 (proved in Appendix D).

*Theorem 5:* The distortion exponent for source coding diversity as described by (61) is

$$\Delta_{\text{SCDIV}} = \max\left[\frac{8\beta p}{4\beta + 3p}, \frac{4\beta p}{4\beta + p}\right]. \tag{62}$$

When $p \leq 4\beta$, MD source coding achieves diversity in the sense that if either channel is bad but the other is good a coarse-grained description of the source can be reconstructed while if both channels are good, a fine-grained description can be reconstructed. Therefore, in this regime, source coding diversity dominates sub-optimal channel coding diversity because it takes advantage of the redundancy between descriptions at the source coding layer.

When $p \geq 4\beta$, however, the max in (62) selects the second term. In this regime, it is more important to maximizes the transmitted rate than protect against fading. Thus source coding diversity degenerates into multiplex channel coding diversity as analyzed in Section IV-D.

In both regimes, optimal channel coding diversity dominates source coding diversity.

## G. Rayleigh Fading AWGN Example

In this section, we evaluate the various distortion exponents on a complex Rayleigh fading additive white Gaussian noise (AWGN) channel. The high SNR approximation for the mutual information on each Rayleigh fading AWGN channel is $F_{e^l}(t) \approx (t/\text{SNR})$, *i.e.*, $p = 1$ in (44) (*e.g.*, see [55], [56] for a discussion of such high SNR expansions).

The resulting distortion exponents are summarized[6] in Tab. II and plotted in Fig. 11. When the processing gain is small (*i.e.*, $\beta \ll 1$), multiplex and optimal channel coding diversity as well as source coding diversity all approach a distortion exponent of $4\beta$, while selection channel coding diversity and no diversity both approach distortion exponents of $2\beta$. Intuitively, this occurs because since bandwidth is scarce, a good system should try to maximize the information communicated by sending different information on each channel. Multiplex coding does this by sending different information on each channel using the same code, optimal channel coding does this by using a different code for each channel, and multiple descriptions coding does this by sending different source descriptions on each channel. Since neither selection diversity nor no diversity provide any multiplexing gain (in the sense of [48]) both of these systems achieve the same sub-optimal distortion exponent.

When the processing gain is large (*i.e.*, $\beta \gg 1$), selection and optimal channel coding diversity as well as source coding diversity all approach a distortion exponent of 2, while systems with multiplex channel coding diversity or no diversity achieve a smaller distortion exponent of 1. Intuitively, this occurs because, since bandwidth is plentiful, even one good channel provides plenty of rate to send a satisfactory description of the source. Thus good systems should try to maximize robustness by being able to decode even if one channel fails completely.

---

[6]The distortion exponents in this paper are slightly different than in [49] due to different definitions of the processing gain as described in Section II-D.1.

At both extremes of processing gain, the best distortion exponent can be achieved either by exploiting diversity at the physical layer via parallel channel coding or at the application layer via multiple description coding. In some sense, this suggests that both physical layer and application layer systems are flexible enough to incorporate the main principles of diversity for continuous channels. Other sub-optimal schemes such as selection channel coding diversity are less flexible in that they only incorporate a subset of the important principles of diversity and thus approach the best distortion exponent in at most one extreme of processing gain. For all processing gains, however, optimal channel coding diversity is superior to source coding diversity, suggesting that the application layer system is missing something. In Section V, we show that the loss of source coding diversity is essentially caused by separating the process of channel decoding from source decoding.

TABLE II

DISTORTION EXPONENTS.

| System | $\Delta$ |
|---|---|
| No Diversity (Section IV-B) | $2\beta/(2\beta+1)$ |
| Selection Channel Coding Diversity (Section IV-E) | $2\beta/(\beta+1)$ |
| Multiplex Channel Coding Diversity (Section IV-D) | $4\beta/(4\beta+1)$ |
| Optimal Channel Coding Diversity (Section IV-E) | $4\beta/(2\beta+1)$ |
| Source Coding Diversity (Section IV-F) | $\max[8\beta/(4\beta+3), 4\beta/(4\beta+1)]$ |

Fig. 12 shows the average distortion for various systems transmitting over complex Rayleigh fading AWGN channels with $\beta = 1$ where the parameters in the rate optimizations have been numerically computed for each system using the high SNR approximations. As the plot indicates, the difference in performance suggested by the asymptotic results in Tab. II becomes evident even at reasonable SNR. Indeed, as the figure shows, optimal channel coding diversity is always superior to source diversity and achieves an advantage of a few dB at moderate SNR. Source diversity is superior to selection diversity by a similar margin. In contrast, Fig. 10 shows that for on-off channels, source-diversity is always better than optimal channel coding diversity for on-off channels. Evidently, none of the systems considered so far are universally optimal and the best way to achieve diversity depends on the qualitative features of the channel. In the next section, we consider a joint source-channel coding system which we show achieves the benefits of source-diversity for on-off channels and the benefits of optimal channel diversity for continuous state channels.

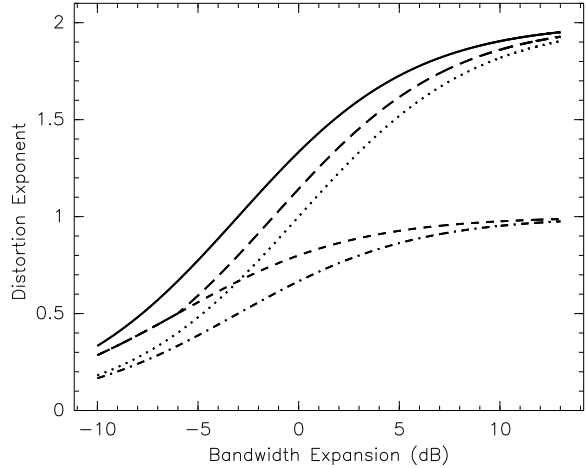

Fig. 11.   Distortion exponents as a function of bandwidth expansion factor $\beta$ in decibels. From top to bottom on the right hand side the curves correspond to optimal channel coding diversity (Section IV-E), source coding diversity (Section IV-F), selection channel coding diversity (Section IV-C), multiplexed channel coding diversity (Section IV-D), and no diversity (Section IV-B).

## V. SOURCE CODING DIVERSITY WITH JOINT DECODING

In this section we consider source coding diversity with a joint decoder that uses the redundancy in both the source coder and channel coder to decode the received signal. Specifically, we define source coding diversity with joint decoding to have encoder and decoder

$$(\mathbf{x}_1, \mathbf{x}_2) = \mathrm{ENC}_{\mathbf{x}_1, \mathbf{x}_2 \leftarrow \mathbf{s}}(\mathbf{s}) = (\mathrm{ENC}_{\mathbf{x}_1 \leftarrow m_1}(\mathrm{ENC}_{m_1 \leftarrow \mathbf{s}}(\mathbf{s})), \mathrm{ENC}_{\mathbf{x}_2 \leftarrow m_2}(\mathrm{ENC}_{m_2 \leftarrow \mathbf{s}}(\mathbf{s}))) \qquad (63a)$$

$$\hat{\mathbf{s}} = \mathrm{DEC}_{\hat{\mathbf{s}} \leftarrow \mathbf{y}_1, \mathbf{y}_2}(\mathbf{y}_1, \mathbf{y}_2) \qquad (63b)$$

where $\mathrm{ENC}_{\mathbf{x}_1 \leftarrow m_1}(\cdot)/\mathrm{ENC}_{\mathbf{x}_1 \leftarrow m_1}(\cdot)$ are single channel encoders (with potentially but not necessarily different codes), $\mathrm{ENC}_{m_1 \leftarrow \mathbf{s}}(\cdot)/\mathrm{ENC}_{m_2 \leftarrow \mathbf{s}}(\cdot)$ are MD source encoders, and $\mathrm{DEC}_{\hat{\mathbf{s}} \leftarrow \mathbf{y}_1, \mathbf{y}_2}(\cdot)$ is a joint source-channel decoder to be described in the sequel.

The motivation for joint source-channel decoding is illustrated by considering the conceptual diagram of an MD quantizer in Fig. 13. Since the two quantization indexes $\mathrm{ENC}_{m_1 \leftarrow \mathbf{s}}(\mathbf{s})$ and $\mathrm{ENC}_{m_2 \leftarrow \mathbf{s}}(\mathbf{s})$ are correlated, the channel decoder should take this correlation into account. For example, if one channel is good and $\mathbf{y}_1$ is accurately decoded to $m_1 = \mathrm{ENC}_{m_1 \leftarrow \mathbf{s}}(\mathbf{s})$ this decreases the number of possible values for $m_2$ and makes decoding $\mathbf{y}_2$ easier.

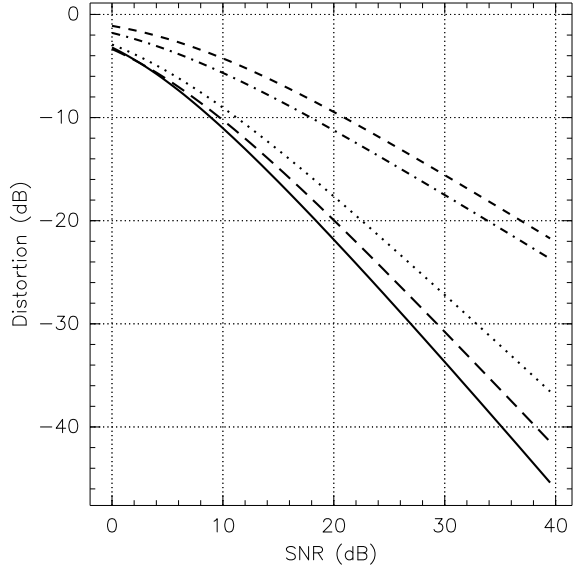

Fig. 12. Average distortion performance on a complex Rayleigh fading additive white Gaussian noise channel with processing gain $\beta = 1$. From top to bottom on the right hand side the curves correspond to no diversity (Section IV-B), multiplexed channel coding diversity (Section IV-D), selection channel coding diversity (Section IV-C), source coding diversity (Section IV-F), and optimal channel coding diversity (Section IV-E).

We show that a joint decoder that exploits this correlation can enlarge the region where both $m_1$ and $m_2$ are successfully decoded. Specifically, with separate decoding, both descriptions are decoded when both $I(\mathbf{x}_1; \mathbf{y}_1)$ and $I(\mathbf{x}_2; \mathbf{y}_2)$ exceed some rate threshold $R_T$, which is denoted as region III in Fig. 3. A joint decoder, however, also recovers both descriptions in region II yielding the decoding regions shown in Fig. 14. With these enlarged decoding regions, we show that source coding diversity with joint source-channel decoding achieves the same performance as optimal channel coding diversity for continuous channels in addition to providing the benefits of source coding diversity for on-off channels.

## A. System Description

Next we describe one way to implement the architecture in (63) using an information theoretic formulation and random coding arguments.

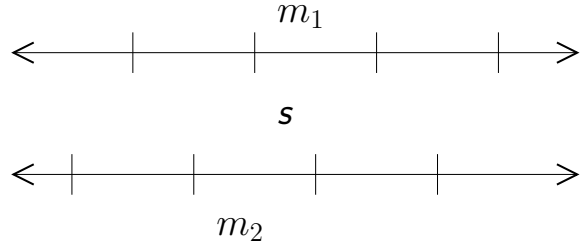

Fig. 13. Conceptual diagram of an MD quantizer. The source $s$ is mapped to the quantizer bins labeled $m_1 = \text{ENC}_{m_1 \leftarrow s}(s)$ and $m_2 = \text{ENC}_{m_2 \leftarrow s}(s)$. Since only overlapping pairs of indexes are legal quantization values, if a receiver accurately decodes $m_1$ from the channel output $\mathbf{y}_1$, then there are only two possible values for $m_2$ in decoding a second channel output $\mathbf{y}_2$.

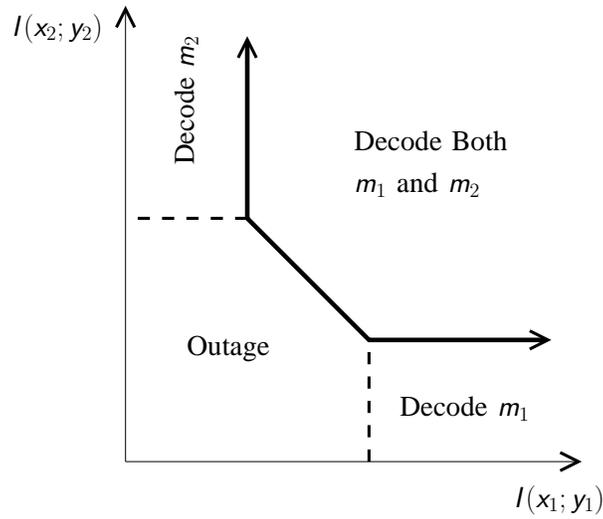

Fig. 14. Decoding regions for a joint source-channel decoder.

*1) Source Encoding:* Choose a test-channel distribution $p_{\hat{s}_1, \hat{s}_2 | s}(\hat{s}_1, \hat{s}_2 | s)$ with the marginal distributions

$$p_{\hat{s}_i}(\hat{s}_i) = \sum_{s, \hat{s}_{3-i}} p_{\hat{s}_1, \hat{s}_2 | s}(\hat{s}_1, \hat{s}_2 | s) p_s(s), \qquad \text{for } i \in \{1, 2\}. \tag{64}$$

Create a pair of rate $R$ random source codebooks, $\mathcal{C}_1$ and $\mathcal{C}_2$ by randomly generating $\exp n_s R$ sequences of length $n_s$ according to the i.i.d. test-channel distributions $p_{\hat{s}_i}(\hat{s}_i)$. To encode a source, find a pair of codewords $\hat{\mathbf{s}}_1 \in \mathcal{C}_1$, $\hat{\mathbf{s}}_2 \in \mathcal{C}_2$, such that the triple $(\hat{\mathbf{s}}_1, \hat{\mathbf{s}}_2, \mathbf{s})$ is strongly typical. According to [28], encoding will succeed with probability approaching one if[7]

$$R > I(\hat{s}_1; s) \tag{65a}$$

$$R > I(\hat{s}_2; s) \tag{65b}$$

$$2R > I(s; \hat{s}_1 \hat{s}_2) + I(\hat{s}_1; \hat{s}_2). \tag{65c}$$

*2) Channel Encoding:* For each channel, generate a rate $R$ random codebook, $\mathcal{C}_i$, by randomly selecting $\exp(n_s R)$ sequences (or equivalently $\exp(n_c R/\beta)$ sequences) of length $n_c$ according to the i.i.d. distribution $p_x(x)$. Encode the source codeword in the $i$th row of $\mathcal{C}_j$ by mapping it to the $i$th channel codeword in $\mathcal{C}_j$.

*3) Joint Decoding:* Denote the output of channel $j$ as $\mathbf{y}_j$ for $j \in \{1,2\}$. To decode, create the lists $\mathcal{L}_1$ and $\mathcal{L}_2$, by finding all channel codewords, $\mathbf{x}_j \in \mathcal{C}_j$, such that the pair $(\mathbf{x}_j, \mathbf{y}_j)$ is typical with respect to the distribution $p_{y_j, x_j | a_j}(y, x | a_j)$. Next search for a unique pair of codewords $(\mathbf{x}_1, \mathbf{x}_2)$ with $\mathbf{x}_1 \in \mathcal{C}_1$ and $\mathbf{x}_2 \in \mathcal{C}_2$ such that the corresponding source codewords $(\hat{s}_1, \hat{s}_2)$ are typical with respect to the distribution $p_{\hat{s}_1, \hat{s}_2}(\hat{s}_1, \hat{s}_2)$. If a unique pair is found, output the resulting source reconstructions. Otherwise declare a decoding error.

*4) Probability Of Error:* The following theorem provides an achievable rate for source coding diversity with joint decoding.

*Theorem 6:* Joint decoding will succeed with probability approaching one if

$$\max\left[0, R - \beta \cdot I(x_1; y_1)\right] + \max\left[0, R - \beta \cdot I(x_2; y_2)\right] \leq I(\hat{s}_1; \hat{s}_2). \tag{66}$$

*Proof:* Decoding can fail if either the correct pair of source codewords are not typical or if an incorrect pair of source codewords are typical. According to the law of large numbers the probability of the former event tends to zero as the block length increases. Therefore, the union bound implies that if the probability of the latter tends to zero, then the total probability of a decoding error also tends to zero.

The probability that an incorrect pair of channel codewords is typical according to $p_{y_j, x_j | a_j}(y, x | a_j)$ is roughly $\exp{-n_c I(x_j; y_j)}$. Since there are $\exp n_s R$ possible codewords for each channel, the expected list sizes are

$$|\mathcal{L}_j| = 1 + \exp\left[n_s R - n_c I(x_j; y_j)\right] + \epsilon \tag{67}$$

---

[7]Note that [28] also includes a term $\hat{s}_0$ which can be ignored (*i.e.*, $\hat{s}_0$ can be set to null or set to a constant such as 0) for our purposes.

where the "1" corresponds to the correct channel codeword and $\epsilon$ denotes a quantity which goes to 0. Using standard arguments it is possible to show that the actual list sizes will be close to the expected list size with probability approaching one.

The probability that an incorrect pair of source codewords, $(\hat{\mathbf{s}}_1, \hat{\mathbf{s}}_2)$ corresponding to the channel codeword pair $(\mathbf{x}_1, \mathbf{x}_2)$ with $\mathbf{x}_j \in \mathcal{C}_j$ is typical is roughly $\exp -n_s I(\hat{s}_1; \hat{s}_2)$. Multiplying this probability by the number of incorrect pairs yields the expected number of incorrect codewords which are nonetheless typical:

$$\exp\left\{-n_s I(\hat{s}_1; \hat{s}_2) + \max\left[0, n_s R - n_c I(x_1; y_1)\right] + \max\left[0, n_s R - n_c I(x_2; y_2)\right]\right\}. \tag{68}$$

Therefore, after dividing through by $n_s$ and recalling that the processing gain is defined as $\beta = n_c/n_s$, we conclude that decoding succeeds provided that (66) holds. ∎

## B. Performance

In order to analyze performance, we must first choose a distribution for the source and channel codebooks. Naturally, we choose the capacity optimizing input distribution for each channel codebook $\mathcal{C}_j$. For the source codebook distribution we use a simpler form of the additive noise test-channel in [28]:

$$\hat{s}_j = s + n_j \tag{69}$$

where $(n_1, n_2)$ is a pair of zero-mean, variance $\sigma^2$, Gaussian random variables independent of $s$ and each other. For this distribution, the distortion when using only description $j$ is $D_j \leq \sigma^2$. When both descriptions are received they can be averaged to yield distortion $D_{1,2} \leq \sigma^2/2$.

*1) Performance on Continuous Channels:* To derive the performance on continuous channels, we must choose $\sigma^2$ as a function of the channel parameters. The choice of $\sigma^2$ determines the rate and hence also the probability of outage and the distortion exponent. Our goal is to show that source coding diversity with joint decoding achieves the same distortion exponent as optimal channel coding diversity. Hence instead of solving an optimization problem to determine $\sigma^2$, we make an educated guess inspired by (101) to choose[8]

$$\sigma^2 = \mathrm{SNR}^{\frac{-4p\beta}{p+2\beta}}. \tag{70}$$

---

[8]Technically, it would be better to choose $\sigma^2$ to be proportional to the right hand side of (70) with a complicated proportionality constant. Since distortion exponent analysis essentially ignores constant factors, however, we ignore this refinement to simplify the exposition.

*Theorem 7:* The distortion exponent for source coding diversity with joint decoding is at least as good as that for optimal channel coding diversity:

$$\Delta_{\text{SCDIV}-\text{JD}} \geq \Delta_{\text{OPT}-\text{CCDIV}}. \tag{71}$$

Note that to achieve the distortion exponent in the previous theorem, the multiple description source redundancy is used in two qualitatively different ways. First, the redundancy between $\mathbf{x}_1$ and $\mathbf{x}_2$ is used to recover the two source descriptions. In this sense, the source coding redundancy acts like channel coding redundancy in providing robustness to noise.

Next, the redundancy between $\hat{\mathbf{s}}_1$ and $\hat{\mathbf{s}}_2$ is used to produce a better source reconstruction by combining the two descriptions. For example, [7] describes a system where the quantization noise for each description is independent of the source and so by averaging the two descriptions, the quantization noise power can be reduced by half. Regardless of how the two descriptions are combined into a higher resolution description, however, the key benefit of joint source-channel decoding is that it can gain the maximum benefit of the redundancy required by multiple description coding both at the channel decoding stage and the source decoding stage.

## VI. Concluding Remarks

We considered various architectures to minimize the average distortion in transmitting a source over independent parallel channels. Conceptually, we view the overall channel quality encountered by a system as a two-dimensional random variable where the two axes correspond to the Shannon mutual information for each channel. As illustrated in Fig. 3, the different architectures considered essentially correspond to systems which perform well when the channel quality is in a certain part of this two-dimensional mutual information plane. Thus minimizing the distortion for a given channel model corresponds to choosing an architecture matched to the shape of the overall channel mutual information distribution.

For on-off channel models, where a channel either fails completely or functions normally, the overall channel mutual information takes values on the Cartesian product of a finite set. This shape is well matched to source coding diversity, *i.e.*, MD source coding and independent channel coding, that exploits diversity at the application layer. Specifically, in the high SNR regime, it is essential that both channels carry redundant information so that if one channel fails the signal can still be decoded from the surviving channel. This forces channel coding diversity to use complete redundancy, and so the distortion when both channels are on is the same as when only one channel is on. In contrast, source coding diversity can use only partial redundancy by sending slightly different signals on each channel. When both channels are on, the differences in the two received descriptions lead to a higher resolution reconstruction and lower distortion. Therefore, source coding diversity achieves substantially better performance than channel coding diversity as illustrated in Fig. 10.

In contrast, for fading, shadowing, and similar effects, the overall channel mutual information takes on a continuous range of values. This shape is better suited to optimal channel coding diversity that exploits at the physical layer. Specifically, in the high SNR regime, optimal channel coding diversity takes advantage of redundancy between the information transmitted across each channel while source coding diversity with separate decoding cannot. As one of our main results, we showed that for such channels the average distortion asymptotically behaves as $\mathrm{SNR}^{-\Delta}$. In particular, we calculated the distortion exponent $\Delta$ for various architectures and showed that the distortion exponent for optimal channel coding diversity is strictly better than for source coding diversity.

Finally, we demonstrated that there is no inherent flaw in source coding diversity on continuous channels. Instead, the inferior distortion exponent of source coding diversity is due to the sub-optimality of separate source and channel decoding. If joint source-channel decoding is allowed, source coding diversity achieves the same distortion exponent as optimal channel coding diversity. Thus, for the non-ergodic channels considered in this paper, Shannon's source-channel separation theorem fails,[9] and the best overall performance is achieved by a joint source-channel architecture using multiple description coding.

While this paper explores a variety of architectures, many aspects of the detailed design, analysis and implementation of such systems remain to be addressed. On the information theory side, determining the best possible average distortion, or at least lower bounds to the best distortion, would be a valuable step. Similarly, determining the performance for architectures using broadcast channel codes combined with successive refinement source codes, hybrid digital-analog codes, or other joint source-channel architectures would be interesting. Also, determining second-order performance metrics beyond the distortion exponent would be useful in designing practical systems. Some issues of interest in signal processing and communication theory include developing practical codes achieving the theoretical advantages of joint source-channel decoding, generalizing the results in this paper to sources with memory or correlated channels (*e.g.*, as found in multiple antenna systems), and studying the effect of imperfect channel state information at the receiver. Finally, a wide array of similar questions arise in a variety of network problems such as relay channels, multi-hop channels, and interference channels. For network scenarios, both the number of possible architectures as well as the advantages of sophisticated systems will be larger.

---

[9]We believe that the main value of Shannon's original source-channel separation theorem was in showing that bits are a sufficient currency between source and channel coding systems. Thus even though the system in Section V has separate encoding and only the decoding is performed jointly, we say that the separation theorem breaks down because exchanging bits is no longer sufficient. Specifically, such a joint decoding system would need to pass lists, log-likelihood ratios, or similar information from the channel coding layer to the source coding layer.

## ACKNOWLEDGMENT

The authors would like to thank Dr. Susie Wee and Dr. Mitchell Trott of Hewlett-Packard Laboratories for many valuable discussions.

## APPENDIX

### A. Distortion Exponent For Selection Channel Coding Diversity

*Proof of Theorem 2:* The minimum expected distortion for such a scheme is computed as follows:

$$E[D] = \min_D \Pr\left\{\max\left[I(x_1; y_1), I(x_2; y_2)\right] < R(D)\right\}$$

$$+ \Pr\left\{\max\left[I(x1; y_1), I(x_2; y_2)\right] \geq R(D)\right\} \cdot D \tag{72}$$

$$= \min_D F_{e'}(\exp R(D))^2 + \left[1 - F_{e'}(\exp R(D))^2\right] \cdot D \tag{73}$$

$$\approx \min_D c^2 D^{\frac{-p}{\beta}} \mathrm{SNR}^{-2p} + \left(1 - c^2 D^{\frac{-p}{\beta}} \mathrm{SNR}^{-2p}\right) \cdot D \tag{74}$$

$$\approx \min_D c^2 D^{\frac{-p}{\beta}} \mathrm{SNR}^{-2p} + D. \tag{75}$$

Differentiating and setting equal to zero yields

$$D^* = \mathrm{SNR}^{\frac{-2p\beta}{p+\beta}} \cdot \left(\frac{\beta}{pc}\right)^{\frac{-p\beta}{p+\beta}} \tag{76}$$

and thus

$$E[D] \approx C_{\mathrm{SEL-CCDIV}} \mathrm{SNR}^{\frac{-2p\beta}{p+\beta}} \tag{77}$$

where $C_{\mathrm{SEL-CCDIV}}$ is a constant independent of SNR. ∎

## B. Distortion Exponent For Multiplexed Channel Coding Diversity

*Proof of Theorem 3:* The minimum expected distortion for such a scheme is computed as follows:

$$E[D] = \min_D \Pr\left\{\min\left[I(x_1; y_1), I(x_2; y_2)\right] < R(D)\right\}$$

$$+ \Pr\left\{\min\left[I(x1; y_1), I(x_2; y_2)\right] \geq R(D)\right\} \cdot D \tag{78}$$

$$= \min_D 2F_{e^l}(\exp[R(D)/2]) - F_{e^l}(\exp[R(D)/2])^2 + [1 - F_{e^l}(\exp[R(D)/2])]^2 \cdot D \tag{79}$$

$$\approx \min_D 2cD^{\frac{-p}{4\beta}}\mathrm{SNR}^{-p} - c^2 D^{\frac{-p}{2\beta}}\mathrm{SNR}^{-2p} + \left(1 - cD^{\frac{-p}{4\beta}}\mathrm{SNR}^{-p}\right)^2 \cdot D \tag{80}$$

$$\approx \min_D 2cD^{\frac{-p}{4\beta}}\mathrm{SNR}^{-p} + D \tag{81}$$

Differentiating and setting equal to zero yields the optimizing distortion

$$D^* = \mathrm{SNR}^{\frac{-4p\beta}{p+4\beta}} \cdot \left(\frac{2\beta}{pc}\right)^{\frac{-4\beta}{p+4\beta}} \tag{82}$$

and thus

$$E[D] \approx C_{\mathrm{MPX-CCDIV}}\mathrm{SNR}^{\frac{-4p\beta}{p+4\beta}} \tag{83}$$

where $C_{\mathrm{MPX-CCDIV}}$ is a constant independent of SNR. ∎

## C. Distortion Exponent for Optimal Channel Coding Diversity

Before proving Theorem 4 we require the following lemma characterizing the mutual information for the parallel channel in terms of probability distribution for each sub-channel.

*Lemma 1:* Let

$$I(\mathbf{x}; \mathbf{y}) = I(x_1; y_1) + I(x_2; y_2)$$

be the mutual information for the total channel and assume that the density and distribution for each sub-channel is given by. (43) and (44) If we define the cumulative distribution function for $\exp I(\mathbf{x}; \mathbf{y})$ as $F_{e^{l_0+l_1}}(t)$ then

$$F_{e^{l_0+l_1}}(t) \approx pc^2 \left(\frac{t}{\mathrm{SNR}^2}\right)^p \left(\ln t - \frac{1}{p}\right) \tag{84}$$

in the sense that the ratio of these quantities goes to 1 as $\mathrm{SNR} \to \infty$.

*Proof:* Note that for any random variable, $a$ with density $f_a(t)$, we have

$$f_{e^a}(t) = f_a(\ln t)/t \text{ and } f_a(t) = f_{e^a}(e^t) \cdot e^t. \tag{85}$$

Therefore we can obtain the desired result by computing the pdf, $f_{I_0+I_1}(t)$, via convolution, and applying (85):

$$f_{I_0+I_1}(t) = \int_0^t f_I(\tau) \cdot f_I(t-\tau) d\tau \tag{86}$$

$$= \int_0^t e^\tau f_{e^I}(e^\tau) \cdot e^{t-\tau} f_{e^I}(e^{t-\tau}) d\tau \tag{87}$$

$$\approx \int_0^t cp \frac{e^{p\tau}}{\text{SNR}^p} \cdot cp \frac{e^{p(t-\tau)}}{\text{SNR}^p} d\tau \tag{88}$$

$$= c^2 p^2 \frac{e^{pt}}{\text{SNR}^{2p}} \int_0^t d\tau \tag{89}$$

$$= c^2 p^2 t \frac{e^{pt}}{\text{SNR}^{2p}} \tag{90}$$

$$F_{e^{I_0+I_1}}(t) = \int_{-\infty}^t f_{e^{I_0+I_1}}(\tau) d\tau \tag{91}$$

$$= \int_{-\infty}^t \frac{f_{I_0+I_1}(\ln\tau)}{\tau} d\tau \tag{92}$$

$$\approx \frac{c^2 p^2}{\text{SNR}^{2p}} \int_1^t \tau^{p-1} \cdot \ln\tau \, d\tau \tag{93}$$

$$= \frac{c^2 p^2}{\text{SNR}^{2p}} \cdot \left( \frac{t^p \ln t}{p} - \frac{t^p}{p^2} + \frac{1}{p^2} \right) \tag{94}$$

$$\approx pc^2 \left( \frac{t}{\text{SNR}^2} \right)^p \left( \ln t - \frac{1}{p} \right) \tag{95}$$

where (88) follows from the high SNR approximation in (43), (93) follows from substituting (90) into (92) and noting that since $I(\mathbf{x}; \mathbf{y})$ is positive then $f_{I_0+I_1}(\ln t)$ is non-zero only for $t > 1$, and the final line follows from noting that the last parenthesized term in (94) is negligible at high SNR. ∎

*Proof of Theorem 4:* To compute the minimum average distortion we have

$$E[D] = \min_D \Pr[I(x_1; y_1) + I(x_2; y_2) < R(D)] + \{1 - \Pr[I(x_1; y_1) + I(x_2; y_2) < R(D)]\} \cdot D \tag{96}$$

$$= \min_D F_{e^{I_0+I_1}}(\exp R(D)) + [1 - F_{e^{I_0+I_1}}(R(D))] \cdot D \tag{97}$$

$$\approx \min_D pc^2 \frac{D^{\frac{-p}{2\beta}}}{\text{SNR}^{2p}} \left( -\beta \ln D - \frac{1}{p} + \frac{D^{\frac{p}{2\beta}}}{p} \right) + \left[ 1 - pc^2 \frac{D^{\frac{-p}{2\beta}}}{\text{SNR}^{2p}} \left( -\beta \ln D - \frac{1}{p} + \frac{D^{\frac{p}{2\beta}}}{p} \right) \right] \cdot D \tag{98}$$

$$\approx \min_D pc^2 \frac{D^{\frac{-p}{2\beta}}}{\text{SNR}^{2p}} \left( -\beta \ln D - \frac{1}{p} + \frac{D^{\frac{p}{2\beta}}}{p} \right) + D. \tag{99}$$

By noting that the parenthesized term in (99) is between $1/p$ and $(1 + e^{-1})/p$ when $D < \exp -(1/p\beta)$, we obtain

$$\min_D c^2 \frac{D^{\frac{-p}{2\beta}}}{\text{SNR}^{2p}} + D \lessapprox E[D] \lessapprox \min_D (1 + e^{-1})c^2 \frac{D^{\frac{-p}{2\beta}}}{\text{SNR}^{2p}} + D. \tag{100}$$

Differentiating the lower bound and setting equal to zero yields the optimizing distortion

$$D^* = \text{SNR}^{\frac{-4p\beta}{p+2\beta}} \cdot \left(\frac{c^2 p}{2\beta}\right)^{\frac{-2\beta}{p+2\beta}}. \tag{101}$$

Substituting (101) into (100) yields

$$C_{\text{LB}} \cdot \text{SNR}^{\frac{-4p\beta}{p+2\beta}} \lessapprox E[D] \lessapprox C_{\text{UB}} \cdot \text{SNR}^{\frac{-4p\beta}{p+2\beta}} \tag{102}$$

where $C_{\text{LB}}$ and $C_{\text{UB}}$ are terms independent of SNR. Hence we conclude that the distortion exponent is

$$\Delta_{\text{OPT−CCDIV}} = (4p\beta)/(p + 2\beta). \tag{103}$$

∎

### D. Distortion Exponent for Source Coding Diversity

*Proof of Theorem 5:* For small $D_0$ and $D_1$, the average distortion is

$$E[D] = \min_{D_0,D_1} \Pr[I(x_1;x_2) < R_{\text{md}}(D_0,D_1)/2]^2$$

$$+ 2\Pr[I(x_1;x_2) < R_{\text{md}}(D_0,D_1)/2] \cdot \Pr[I(x_1;x_2) \geq R_{\text{md}}(D_0,D_1)/2] \cdot D_1$$

$$+ \Pr[I(x_1;x_2) \geq R_{\text{md}}(D_0,D_1)/2]^2 \cdot D_0 \tag{104}$$

$$= \min_{D_0,D_1} F_{e^I}(\exp R_{\text{md}}(D_0,D_1)/2)^2 + 2 \cdot F_{e^I}(\exp R_{\text{md}}(D_0,D_1)/2) \cdot [1 - F_{e^I}(\exp R_{\text{md}}(D_0,D_1)/2)] \cdot D_1$$

$$+ [1 - F_{e^I}(\exp R_{\text{md}}(D_0,D_1)/2)]^2 \cdot D_0 \tag{105}$$

$$\approx \min_{D_0,D_1} \frac{c^2}{\text{SNR}^{2p}} \exp\{p \cdot R_{\text{md}}(D_0,D_1)\}$$

$$+ 2\frac{c}{\text{SNR}^p} \exp\left\{\frac{p}{2} \cdot R_{\text{md}}(D_0,D_1)\right\} \cdot \left[1 - \frac{c}{\text{SNR}^p} \exp\left\{\frac{p}{2} \cdot R_{\text{md}}(D_0,D_1)\right\}\right] \cdot D_1$$

$$+ \left[1 - \frac{c}{\text{SNR}^p} \exp\left\{\frac{p}{2} \cdot R_{\text{md}}(D_0,D_1)\right\}\right]^2 \cdot D_0 \tag{106}$$

$$\approx \min_{D_0,D_1} \frac{c^2}{\text{SNR}^{2p}} \exp\{p \cdot R_{\text{md}}(D_0,D_1)\} + 2\frac{c}{\text{SNR}^p} \exp\left\{\frac{p}{2} \cdot R_{\text{md}}(D_0,D_1)\right\} \cdot D_1 + D_0. \tag{107}$$

Substituting the bounds from (30) into (107) yields

$$E[D] \lessapprox \min_{D_1,D_0} \frac{c^2}{\text{SNR}^{2p}} \left(\frac{1}{2D_1 D_0}\right)^{\frac{p}{2\beta}} + \frac{2c}{\text{SNR}^p} \left(\frac{1}{2D_1 D_0}\right)^{\frac{p}{4\beta}} \cdot D_1 + D_0 \tag{108a}$$

$$E[D] \gtrapprox \min_{D_1,D_0} \frac{c^2}{\text{SNR}^{2p}} \left(\frac{1}{4D_1 D_0}\right)^{\frac{p}{2\beta}} + \frac{2c}{\text{SNR}^p} \left(\frac{1}{4D_1 D_0}\right)^{\frac{p}{4\beta}} \cdot D_1 + D_0 \tag{108b}$$

where (108b) requires $D_0 \to 0$ and (108a) also requires $D_1 \to 0$.

When $p \geq 4\beta$ then (108) increases as $D_1$ becomes small. Hence in this regime the optimal choice for $D_1$ approaches a constant bounded away from zero. If the low distortion formula for the lower bound is used, then the optimal choice for $D_1$ approaches one. Technically, however, for $D_1 \geq 1/2$ the rate required is given by (26b) not (26a), so there is no excess rate in multiple description coding [16], [28] and the optimal $D_1$ for $p \geq 4\beta$ approaches $1/2$ using (26b). In any case, regardless of whether $D_1 = 1/2$ or $D_1 = 1$ or some other intermediate value, when $p \geq 4\beta$, average distortion is minimized by choosing $D_1$ to be large. Thus for $p \geq 4\beta$, the optimal multiple description system essentially degenerates into the channel multiplexing scheme analyzed in Section IV-D and achieves the same distortion exponent (although with a slightly different constant factor term).

When $p < 4\beta$, we can find the optimal value for $D_1$ by differentiating the lower bound with respect to $D_1$ and setting equal to 0 to obtain

$$D_1^* = \left( \frac{4\beta - p}{cp} \right)^{\frac{-4\beta}{4\beta - p}} \cdot \mathrm{SNR}^{\frac{-4\beta}{4\beta + p}} \cdot (4D_0)^{-1 + \frac{4\beta}{4\beta + p}}, \, p < 4\beta. \tag{109}$$

For the case when $p < 4\beta$, substituting (109) into (108b) yields

$$E[D] \gtrapprox C \cdot D_0^{\frac{-2p}{4\beta + p}} \cdot \mathrm{SNR}^{\frac{-8p\beta}{4\beta + p}} + D_0 \text{ for } p < 4\beta \tag{110}$$

where $C$ is a constant independent of SNR and $D_0$. Differentiating with respect to $D_0$ and setting the result equal to zero yields the optimal value for $D_0$:

$$D_0^* = \begin{cases} C' \cdot \mathrm{SNR}^{\frac{-8\beta p}{4\beta + 3p}}, & p < 4\beta \\ C'' \cdot \mathrm{SNR}^{\frac{-4\beta p}{4\beta + p}}, & p \geq 4\beta \end{cases} \tag{111}$$

from which we conclude

$$C_{\mathrm{LB}} \cdot \mathrm{SNR}^{- \max \left[ \frac{8\beta p}{4\beta + 3p}, \frac{4\beta p}{4\beta + p} \right]} \lessapprox E[D] \lessapprox C_{\mathrm{UB}} \cdot \mathrm{SNR}^{- \max \left[ \frac{8\beta p}{4\beta + 3p}, \frac{4\beta p}{4\beta + p} \right]} \tag{112}$$

where the max occurs since multiple description coding essentially degenerates into channel multiplexing with a better constant factor when $p \geq 4\beta$. ∎

### E. Distortion Exponent for Source Coding Diversity with Joint Decoding

Computing the exact rates required to guarantee successful encoding in (65) is generally difficult, thus we focus on the high resolution limit in the following Lemma.

*Lemma 2:* Let $s$ be a source with finite variance and finite entropy power. Then in the high resolution limit, choosing

$$R > h(s) - (1/2) \log 2\pi e \sigma^2 \tag{113}$$

asymptotically satisfies (65) and guarantees successful encoding.

*Proof:* Proving the claim requires showing that

$$\lim_{D_j \to 0} I(s; \hat{s}_j) - \left[ h(s) - \frac{1}{2} \log 2\pi e\sigma^2 \right] = 0 \quad \text{for } j \in \{0, 1\} \tag{114}$$

and

$$\lim_{D_j \to 0} I(s; \hat{s}_1 \hat{s}_2) + I(\hat{s}_1; \hat{s}_2) - 2 \left[ h(s) - \frac{1}{2} \log 2\pi e\sigma^2 \right] = 0. \tag{115}$$

The former follows from the fact that the Shannon Lower Bound is asymptotically tight [50]. In the interest of completeness, however, we define $\Delta R$ as left hand side of (114) and summarize the argument showing that it goes to zero:

$$\Delta R \stackrel{\Delta}{=} \lim_{D_j \to 0} I(s; \hat{s}_j) - \left[ h(s) - \frac{1}{2} \log 2\pi e\sigma^2 \right] \tag{116}$$

$$= \lim_{D_j \to 0} h(s + n_j) - h(s + n_j | s) - \left[ h(s) - \frac{1}{2} \log 2\pi e\sigma^2 \right] \tag{117}$$

$$= \lim_{D_j \to 0} h(s + n_j) - h(n_j) - \left[ h(s) - \frac{1}{2} \log 2\pi e\sigma^2 \right] \tag{118}$$

$$= \lim_{D_j \to 0} h(s + n_j) - h(s) \tag{119}$$

$$= 0. \tag{120}$$

Equations (117) and (118) follow from the choice of the conditional distribution $\hat{s}_j = s + n_j$ where $n_j$ is independent of $s$. The key step in going from (119) to (120) is the "continuity" property of differential entropy [50, Theorem 1] which is the main tool in obtaining many high-resolution source coding results.

A similar chain of equalities establishes (115). Specifically, if we define the right hand side of (115) as $\Delta 2R$ then we obtain

$$\Delta 2R \stackrel{\Delta}{=} \lim_{D_j \to 0} I(s; \hat{s}_1 \hat{s}_2) + I(\hat{s}_1; \hat{s}_2) - 2 \left[ h(s) - \frac{1}{2} \log 2\pi e\sigma^2 \right] \tag{121}$$

$$= \lim_{D_j \to 0} h(\hat{s}_1 \hat{s}_2) - h(\hat{s}_1 \hat{s}_2 | s) + h(\hat{s}_1) - h(\hat{s}_1 | \hat{s}_2) - 2 \left[ h(s) - \frac{1}{2} \log 2\pi e\sigma^2 \right] \tag{122}$$

$$= \lim_{D_j \to 0} h(\hat{s}_1) + h(\hat{s}_2) - h(\hat{s}_1 | s) - h(\hat{s}_2 | s) - 2 \left[ h(s) - \frac{1}{2} \log 2\pi e\sigma^2 \right] \tag{123}$$

$$= \lim_{D_j \to 0} 2 \cdot \Delta R \tag{124}$$

$$= 0 \tag{125}$$

where (124) follows by noting that (123) is simply twice (116), and hence (125) follows from (120). ∎

In the sequel, we require the following Lemma which states that, in the high resolution limit, the two descriptions, $\hat{s}_1$ and $\hat{s}_2$, only differ in half a bit per sample. This close relationship between the two descriptions enables the joint decoder to approach the performance of parallel channel coding with a single description.

*Lemma 3:* If the rate is chosen according to (113), specifically, if the difference between the two sides is $\epsilon$, then

$$\lim_{D_j \to 0} I(\hat{s}_1; \hat{s}_2) - R \geq -\frac{1}{2} \log 2 - \epsilon. \tag{126}$$

*Proof:* We have the following chain of inequalities:

$$\lim_{D_j \to 0} I(\hat{s}_1; \hat{s}_2) - R = \lim_{D_j \to 0} h(s + n_1) - h(s + n_1 | s + n_2) - R \tag{127}$$

$$= \lim_{D_j \to 0} h(s + n_1) - h(n_1 - n_2 | s + n_2) - R \tag{128}$$

$$\geq \lim_{D_j \to 0} h(s + n_1) - h(n_1 - n_2) - R \tag{129}$$

$$= \lim_{D_j \to 0} h(s + n_1) - \frac{1}{2} \log 4\pi e \sigma^2 - R \tag{130}$$

$$= \lim_{D_j \to 0} h(s + n_1) - h(n_1) - \frac{1}{2} \log 2 - R \tag{131}$$

$$= \lim_{D_j \to 0} I(\hat{s}_1; s) - R - \frac{1}{2} \log 2 \tag{132}$$

$$= \lim_{D_j \to 0} I(\hat{s}_1; s) - \left[ h(s) - \frac{1}{2} \log 2\pi e + \epsilon \right] - \frac{1}{2} \log 2 \tag{133}$$

$$= \lim_{D_j \to 0} \Delta R - \frac{1}{2} \log 2 - \epsilon \tag{134}$$

$$= -\frac{1}{2} \log 2 - \epsilon. \tag{135}$$

Most of the arguments follow from well-known properties of mutual information and entropy. Equation (135) follows from Lemma 2.

∎

*Proof of Theorem 7:* If we choose $\sigma^2$ as in (70), the expected distortion is at most the distortion when both descriptions are successfully decoded times the probability that both descriptions are not decoded. Hence, applying Theorem 6 yields

$$E[D] \leq \frac{\sigma^2}{2} \cdot \Pr[\mathcal{E}] + \Pr[\mathcal{E}^c] \tag{136}$$

where $\mathcal{E}$ denotes the event that both descriptions can be decoded as defined in (66) and $\mathcal{E}^c$ is the complement of $\mathcal{E}$. Note that since $\Pr[\mathcal{E}] \leq 1$, the first term on the right hand side of (136) is proportional to $\mathrm{SNR}^{-\Delta_{\mathrm{OPT-CCDIV}}}$ by construction due to our choice of $\sigma^2$ in (70). Therefore, to prove the Theorem, we need to bound the second term, $\Pr[\mathcal{E}^c]$.

If we let $\mathcal{E}[i,j]$ (with $i,j \in \{1,2\}$) denote the event that the first $\max$ operation in $\mathcal{E}$ returns the $i$th argument while the second $\max$ operation in $\mathcal{E}$ returns the $j$th argument, then we can express the second term in (136) as

$$\Pr[\mathcal{E}^c] = \Pr[\mathcal{E}^c \cap \mathcal{E}[1,1]|\mathcal{E}[1,1]] \Pr[\mathcal{E}[1,1]] + \Pr[\mathcal{E}^c \cap \mathcal{E}[1,2]|\mathcal{E}[1,2]] \Pr[\mathcal{E}[1,2]]$$
$$+ \Pr[\mathcal{E}^c \cap \mathcal{E}[2,1]|\mathcal{E}[2,1]] \Pr[\mathcal{E}[2,1]] + \Pr[\mathcal{E}^c \cap \mathcal{E}[2,2]|\mathcal{E}[2,2]] \Pr[\mathcal{E}[2,2]]. \quad (137)$$

To prove the theorem, it is sufficient to show that for every $\epsilon > 0$, there exists a constant $c_{i,j}$ such that

$$\Pr[\mathcal{E}^c|\mathcal{E}[i,j]] \Pr[\mathcal{E}[i,j]] \leq c_{i,j} \cdot \mathrm{SNR}^{\epsilon - \Delta_{\mathrm{OPT-CCDIV}}}$$

for large enough SNR.

Conditioned on $\mathcal{E}[1,1]$, both $I(x_1; y_1) > R/\beta$ and $I(x_2; y_2) > R/\beta$, so both channels are good enough to decode each description separately. Thus $\Pr[\mathcal{E}^c|\mathcal{E}[1,1]] = 0$, and therefore $\Pr[\mathcal{E}^c|\mathcal{E}[1,1]] \Pr[\mathcal{E}[1,1]] = 0$ as well. This takes care of the first term in (137).

Next we consider the second term of (137). Conditioned on $\mathcal{E}[1,2]$, only $I(x_1; y_1) > R/\beta$ while $I(x_2; y_2) < R/\beta$ and only description 1 can be decoded separately. Description 2 can be decoded jointly provided that $I(x_2; y_2) \geq R/\beta - I(\hat{s}_1; \hat{s}_2)/\beta$. By applying Lemma 3, this condition becomes $I(x_2; y_2) > (\log 2)/(2\beta)$ in the high-resolution limit, therefore

$$\Pr[\mathcal{E}^c|\mathcal{E}[1,2]] \Pr[\mathcal{E}[1,2]] \approx \Pr\left[I(x_2; y_2) \leq \frac{\log 2}{2\beta}\right] \cdot \Pr[\mathcal{E}[1,2]] \quad (138)$$

$$\approx c \cdot \left(\frac{2^{\frac{1}{2\beta}}}{\mathrm{SNR}}\right)^p \cdot \Pr[\mathcal{E}[1,2]] \quad (139)$$

$$\leq c \cdot \left(\frac{2^{\frac{1}{2\beta}}}{\mathrm{SNR}}\right)^p \cdot \Pr[I(x_2; y_2) < R/\beta] \quad (140)$$

$$\approx c \cdot \left(\frac{2^{\frac{1}{2\beta}}}{\mathrm{SNR}}\right)^p \cdot c \cdot \left(\frac{\exp \frac{h(s)}{\beta}}{\sigma^{1/\beta}\mathrm{SNR}}\right)^p \quad (141)$$

$$= \mathrm{SNR}^{-2p} \cdot \mathrm{SNR}^{\frac{2p}{p+2\beta}} \cdot c^2 \left(2^{\frac{1}{2\beta}} \exp \frac{h(s)}{\beta}\right)^p \quad (142)$$

$$= \mathrm{SNR}^{\frac{-4p\beta}{p+2\beta}} \cdot c^2 \left(2^{\frac{1}{2\beta}} \exp \frac{h(s)}{\beta}\right)^p \quad (143)$$

where in going from (140) to (141) we replaced $R$ with $h(s) - (1/2)\log 2\pi e \sigma^2$ and recalled that we assumed $\exp[2h(s)] = 2\pi e$ just after (23).

Thus, for some constant $C_{\mathrm{SCDIV-JD}}$, and every $\epsilon > 0$, there exists an SNR large enough such that

$$\Pr[\mathcal{E}^c|\mathcal{E}[1,2]] \Pr[\mathcal{E}[1,2]] \leq \mathrm{SNR}^{\epsilon - \frac{4p\beta}{p+2\beta}} \cdot C_{\mathrm{SCDIV-JD}} \quad (144)$$

and

$$\Pr[\mathcal{E}^c | \mathcal{E}[2,1]] \Pr[\mathcal{E}[2,1]] \leq \text{SNR}^{\epsilon - \frac{4p\beta}{p+2\beta}} \cdot C_{\text{SCDIV-JD}}. \tag{145}$$

A similar analysis works for the third term of (137).

Finally, we consider the last term in (137). Conditioned on $\mathcal{E}[2,2]$, both $I(x_1; y_1) < R/\beta$ and $I(x_2; y_2) < R/\beta$, so neither channels is good enough for separate decoding. Successful joint decoding requires

$$I(x_1; y_1) + I(x_2; y_2) > [2R - I(\hat{s}_1; \hat{s}_2)] / \beta. \tag{146}$$

and therefore

$$\Pr[\mathcal{E}^c \cap \mathcal{E}[2,2] | \mathcal{E}[2,2]] = \Pr[I(x_1; y_1) + I(x_2; y_2) \leq 2R/\beta - I(\hat{s}_1; \hat{s}_2)/\beta] \tag{147}$$

$$\stackrel{<}{\approx} \Pr\left[I(x_1; y_1) + I(x_2; y_2) \leq R/\beta - \frac{\log 2}{2\beta}\right] \tag{148}$$

$$\approx pc^2 \left(\frac{2^{\frac{1}{2\beta}} \exp \frac{h(s)}{2\beta}}{\sigma^{1/\beta} \text{SNR}^2}\right)^p \cdot \left(\frac{h(s)}{2\beta} - \frac{\log 2\pi e\sigma^2}{2\beta} + \frac{\log 2}{2\beta} - \frac{1}{p}\right) \tag{149}$$

$$= \sigma^{-p/\beta} \cdot \text{SNR}^{-2p} \cdot 2^{\frac{1}{2\beta}} \cdot \exp \frac{h(s)}{2\beta} \cdot \left(\frac{h(s)}{2\beta} - \frac{\log 2\pi e\sigma^2}{2\beta} + \frac{\log 2}{2\beta} - \frac{1}{p}\right) \tag{150}$$

$$\approx \text{SNR}^{\frac{-2p^2}{p+2\beta} - 2p^2} \cdot C'_{\text{SCDIV-JD}} \cdot \text{SNR}^{\epsilon} \tag{151}$$

$$= \text{SNR}^{\epsilon - \frac{4p\beta}{p+2\beta}} \cdot C'_{\text{SCDIV-JD}} \tag{152}$$

where (148) follows since Lemma 3 implies

$$2R - I(\hat{s}_1; \hat{s}_2) \leq R - \frac{1}{2}\log 2 + \epsilon, \tag{153}$$

$\epsilon$ is a quantity which can be made arbitrarily small, and $C'_{\text{SCDIV-JD}}$ is some constant independent of SNR.

The above results combined with $\Delta_{\text{OPT-CCDIV}} = 4p\beta/(p+2\beta)$ proves the desired result. ∎